\documentclass[preprint2]{aastex}

\def \eg           {{e.g.}}
\def \etal         {{et~al. }}

\def \kms          {\hbox{km$\,$s$^{-1}$}}
\def \kkms         {\hbox{K$\,$km$\,$s$^{-1}$}}

\def\approxlt{\lower.2em\hbox{$\buildrel < \over \sim$}}
\def\approxgt{\lower.2em\hbox{$\buildrel > \over \sim$}}

\def \ls           {\hbox{L$_{\odot}$}}
\def \ms           {\hbox{M$_{\odot}$}}           

\begin{document}
\title{Molecular Gas and the Modest Star Formation Efficiency \\ 
in the ``Antennae'' Galaxies: Arp~244=NGC~4038/9}

\author{Yu Gao\altaffilmark{1,2,3}, K.Y. Lo\altaffilmark{3,4}, 
S.-W. Lee\altaffilmark{2,4}, \& T.-H. Lee\altaffilmark{4,5}}
 
\altaffiltext{1}{Infrared Processing and Analysis Center, 
Jet Propulsion Laboratory, 
    Caltech 100-22, 770 S. Wilson Ave., Pasadena, CA 91125}
\altaffiltext{2}{Department of Astronomy, University 
of Toronto, 60 St. George Street, Toronto, ON M5S 3H8, CANADA}
\altaffiltext{3}{Laboratory for Astronomical Imaging, Department of Astronomy,
        University of Illinois, 1002 W. Green Street, Urbana, IL 61801}
\altaffiltext{4}{Institute of Astronomy and Astrophysics, Academia Sinica,
        P.O. Box 1-87, Nankang, Taipei, Taiwan}
\altaffiltext{5}{Department of Physics and Astronomy,
University of Calgary, 2500 University Drive, N.W., Calgary, Alberta T2N 1N4,
CANADA}

\authoremail{gao@ipac.caltech.edu, kyl@asiaa.sinica.edu.tw,
swlee@fft.astro.utoronto.ca, thlee@iras.ucalgary.ca}

\received{To appear in ApJ, v548, Feb 10, 2001 issue}
 
\begin{abstract}

We report here a factor of 5.7 higher total CO flux in Arp~244
(the ``Antennae'' galaxies) than that
previously accepted in the literature (thus a total molecular gas 
mass of 1.5$\times$10$^{10}$ \ms), based on 
our fully sampled CO(1-0) observations at the NRAO 12m telescope. Currently,
much of the understanding and modeling of the star formation 
in Arp~244 has been derived from a much lower molecular
gas mass. It is imperative to reconsider all as the high
molecular gas mass might provide
sufficient fuel for ultraluminous extreme starburst 
in Arp~244 once the merging advances to late stage. 

Our observations show that the molecular gas peaks predominately in 
the disk-disk overlap region between the nuclei, similar to the 
far-infrared (FIR) and mid-infrared (MIR) emission. The bulk of 
the molecular gas is forming into stars with a normal 
star formation efficiency (SFE) $L_{\rm IR}/M({\rm H}_2) \approx 
4.2 \ls/\ms$, same as that of giant molecular clouds in the Galactic disk.
Additional supportive evidence is the extremely low fraction of 
the dense molecular gas in Arp~244, revealed by our detections of the 
HCN(1-0) emission, which traces the active star-forming 
gas at density $\approxgt 10^4$ cm$^{-3}$.

Using the high-resolution BIMA plus the NRAO 12m telescope, 
full-synthesis CO(1-0) images,
and our VLA continuum maps at 20cm, we estimate the local SFE 
indicated by the ratio map of the radio continuum to CO(1-0) emission,
down to kpc scale. Remarkably, the local SFE stays
roughly same over the bulk of the molecular
gas distribution. Only some localized regions show the highest
radio-to-CO ratios that we have identified as the sites of the most 
intense starbursts with SFE$\approxgt 20~\ls/\ms$. Here we have assumed that 
the 20cm emission is a fairly good indicator of star formation 
down to kpc scale in Arp~244 because of the well-known correlation
between the FIR and the radio continuum emission. 
These starburst regions are confined exclusively in the dusty patches
seen in the HST optical images near the CO and FIR peaks where 
presumably the violent starbursts are heavily obscured. Nevertheless,
recent large-scale star formation is going on throughout
the system (\eg, concentrations of numerous super-star clusters 
and MIR ``hotspots''), yet the measured level is more suggestive of 
a moderate starburst (SFE$\approxgt 10~\ls/\ms$)
or a weak to normal star formation (SFE$\sim 4~\ls/\ms$), not necessarily
occurring at the high concentrations of the molecular gas reservoir. 
The overall starburst from the bulk of the molecular gas is 
yet to be initiated as most of the gas further condenses into kpc
scale in the final coalescence. 

\end{abstract}
 
\keywords{galaxies: individual (NGC~4038/9) --- galaxies: interactions 
--- galaxies: starburst --- 
infrared: galaxies --- ISM: molecules --- stars: formation}
 
\section{INTRODUCTION}

The ``Antennae'' galaxies (Arp~244, NGC~4038/39, VV~245), the nearest 
IR-luminous and perhaps the youngest prototypical 
galaxy-galaxy merger (the first in Toomre (1977) merger sequence),
certainly reclaimed its fame from the recent 
releases of the HST/WFPC2 images (Whitmore \etal 1999) and the
Chandra X-ray images (Fabbiano \etal 2000), and the observations 
at essentially all available wavelengths from radio to X-ray
(\eg,  Hummel \& van der Hulst 1986; Neff, \& Ulvestad 2000; Hibbard, 
van der Hulst, \& Barnes 2001 in preparation; Bushouse, Telesco, 
\& Werner 1998; Nikola \etal 1998; Mirabel \etal 1998; Vigroux \etal 
1996; Read, Ponman, \& Wolstencroft 1995; 
Fabbiano, Schweizer, \& Mackie 1997). 
The HST images reveal over one thousand bright young star clusters
that are thought to have formed in recent bursts of star formation.
H$\alpha$ imaging also shows the most recent locations of the star-forming 
giant HII regions and their velocity fields as well~(Rubin, Ford, \& 
D'Odorico 1970; Amram \etal 1992; Whitmore \etal 1999). 
Soft X-ray~(Fabbiano \etal 1997) and
radio continuum~(Hummel \& van der Hulst 1986) images 
may hint of previous star formation sites currently seen as 
supernova remnants. In addition, mid-infrared (MIR)~(Mirabel \etal 1998;
Vigroux \etal 1996), far-infrared (FIR)~(Evans, Harper, \& Helou 1997;
R. Evans 1998, private communications; Bushouse \etal 1998) 
and submillimeter (Haas \etal 2000) images indicate that the most 
intense starburst takes place currently in an off-nucleus
region that is inconspicuous at optical wavelengths. The star-forming
molecular gas in Arp~244 is perhaps least understood, however; 
we show in this paper that 
the total molecular gas mass accepted in the literature for the last
decade~(Sanders \& Mirabel 1985; Stanford \etal 1990) has been 
underestimated by nearly a factor of 6.

The importance of gas during galaxy-galaxy merging far exceeds 
its mass proportion, as demonstrated by the
sophisticated numerical simulations~(\eg, Barnes \& Hernquist 1996, 1998). 
At a distance of 20 Mpc ($H_0=75$\kms~Mpc$^{-1}$, \eg, van der Hulst 1979;
Mihos, Bothun, \& Richstone 1993; Mirabel \etal 1998), the total IR 
luminosity (8--1000 $\mu$m, Sanders \& Mirabel 1996) of Arp~244,
measured from the {\it IRAS} four-band fluxes
given in Soifer \etal (1989), is $L_{\rm IR}=6.2\times 10^{10} \ls$. 
Using the most recent remeasured total {\it IRAS} fluxes (as opposed to
simply the ``point source'' in Soifer \etal 1989) in the Revised
Bright Galaxy Sample (D.B. Sanders 2000, private communication;
Sanders, Mazzarella, Kim, \& Surace 2001 in preparation), the 
same IR luminosity is obtained.
Therefore, strictly speaking Arp~244 is not a luminous
infrared galaxy (LIG, $L_{\rm IR}\approxgt 10^{11} \ls$) unless
a Virgocentric flow distance of 29.5 Mpc is used (Sanders \etal 2000, but 
we use 20 Mpc throughout this paper), 
which leads to $L_{\rm FIR}=1.0\times 10^{11} \ls$ and
$L_{\rm IR}=1.3\times 10^{11} \ls$.
A much smaller molecular gas mass of $2.6 \times 10^9$~\ms~ 
(Sanders \& Mirabel 1985),
currently accepted in the literature, leads to a higher global ratio of 
$L_{\rm IR}/M({\rm H}_2) \sim 24~\ls/\ms$. This ratio is often
referred as the star formation efficiency (SFE) since the FIR
emission, tracer of current star formation rate,
has been normalized by the molecular gas mass available to make stars.
With this high yield of young stars per unit molecular gas mass and 
a lower molecular gas content, most of the molecular gas will be depleted
in $\sim 10^8$ years, as noted by Sanders \& Mirabel (1985) and 
Stanford \etal (1990). There would be no chance
for Arp~244 to become an ultraluminous infrared galaxy (ULIG,
$L_{\rm IR}\approxgt 10^{12} \ls$) in the late stage of the merging.
Numerical modeling (Mihos \etal 1993), based on 
this lower gas mass, indeed predicted 
that Arp~244 would not join the rank of ULIGs. 

ULIGs, with the IR luminosity 
comparable to the bolometric luminosity of QSOs, are the most luminous 
galaxies in the local universe. They are believed to be powered mainly
by starbursts (e.g., Smith, Lonsdale, \& Lonsdale 1998; Genzel et al. 1998)
taking place predominantly in the {\it extreme starburst} regions of 
a characteristic size of $\sim 100$ pc and 
$L_{\rm IR} \sim 3\times 10^{11} \ls$ 
(Downes \& Solomon 1998), produced as a result of the merging 
of molecular {\it gas-rich} spiral galaxies (Sanders \& Mirabel 1996). 
Alternatively, the dust-enshrouded active galactic nuclei (AGN) 
may still be responsible for significant 
energy output in some ULIGs (Sanders \etal 1988; Veilleux, 
Sanders, \& Kim 1999; Sanders 1999). Now, for a total molecular
gas mass of $1.5\times 10^{10} \ms$, based on our fully
sampled CO(1-0) observations obtained at the NRAO
\footnote{The National Radio Astronomy Observatory is a 
facility of the National Science Foundation operated under 
cooperative agreement by Associated Universities, Inc.} 12m,
many previous conclusions and
speculations about Arp~244 need to be revised. After all, 
the initial gas content, particularly the molecular gas --- 
the fuel for star formation, will probably be the most important
factor in determining whether a merging pair of spiral galaxies
reaches the peak of ultraluminous extreme starburst phase
since all ULIGs are still gas-rich with $\sim 10^{10}$ \ms~ molecular
gas mass (Solomon \etal 1997), most likely concentrated in the
kpc scale disks/rings around the merging nuclei (Downes \& Solomon 1998; 
Scoville, Yun, \& Bryant 1997;
Sakamoto \etal 1999; cf. Evans, Surace, \& Mazzarella 2000).

With a molecular gas mass comparable to that of ULIGs 
and an early/intermediate merging stage, Arp 244 is perhaps
an example of what an ULIG might have looked like a few hundred
million years ago.
Arp 244 may be a snapshot in the evolution of a typical gas-rich merger
into an ULIG system.
Thus, it is crucial to understand how, where and when the starbursts 
initiate or have occurred in such an on-going merger. This becomes 
especially imperative in Arp~244 given
the advantage of its close-up distance and therefore the better linear
resolution available. Nonetheless, the strongest starburst site revealed
from the Infrared Space Observatory ($ISO$) MIR images
(Vigroux \etal 1996; Mirabel \etal 1998; Xu \etal 2000) seems to 
be offset from the peak emission in all the 60, 100, and 160 $\mu$m 
FIR maps of the limited resolution, obtained
with the Kuiper Airborne 
Observatory (KAO, Evans, Harper, \& Helou 1997; Bushouse \etal 1998).
In order to best locate the sites of the intense star formation, 
high-resolution imaging in the FIR is ultimately required. 

Since there is an excellent correlation between the FIR
and the radio continuum emission (\eg, Helou, Soifer, \& Rowan-Robinson 1985;
Condon \etal 1990; Condon 1992; Xu \etal 1994; Marsh \& Helou 1995),
the FIR emission can thus be approximately scaled according to the radio
continuum emission with the high resolution achievable by 
the Very Large Array (VLA) observations. We have therefore, 
obtained the VLA radio continuum images to compare with our 
high-resolution full-synthesis 
Berkeley-Illinois-Maryland Association (BIMA) plus the 
NRAO 12m CO maps. This is because the star formation occurs within 
giant molecular clouds (GMCs), especially the dense cores and 
starburst can be better characterized by the elevated SFE, i.e., 
the FIR to CO ratio, approximated here by the radio-to-CO ratio. 
Therefore, local SFE across over the merging system can be 
approximately measured using the ratio map
of the radio continuum to the CO emission to locate the most intense 
starburst sites of the highest radio-to-CO ratios.

We here briefly report our various observations in \S 2
detailing the single-dish CO mapping in Arp~244. The
BIMA interferometry CO and the VLA 20cm continuum observations 
will be presented elsewhere. The observational results and analysis
are summarized and compared in \S 3. Section 4 discusses
the starburst properties of Arp~244 and the implications of 
our observations. Finally, we conclude our main results.

\section{OBSERVATIONS}

We obtained the NRAO 12m single-dish CO(1-0) map at Nyquist sampling 
(half-beam spacing $\sim 27.''5$) in order 
to determine the true total CO extent and distribution and help
add in the zero-spacing flux missed from the multifield 
BIMA CO(1-0) data cube. A total of more than
50 positions have been observed in 1998 March and 20 more 
positions in the outer edge of the map were further obtained 
in 1998 April to fully cover the entire CO emission region. 
We have essentially integrated for at least 1.5 hr 
at every position in order to make sure
that not only no further $\approxgt 3\sigma$ level CO emission is detected
in the outmost edge of the map but all spectra obtained at different
positions have roughly comparable RMS noise level as well. 
A total of 73 positions have been 
mapped in the merging disks accumulating more than 110 hr useful
integration time. Additional observations in 1999 June and November
have been conducted remotely in the southeastern (SE) extension of 
the disk overlap region (where the much longer southern tidal
tail starts) and at the tip of the southern tail (Figure~1). This is
to further integrate down the noise level to clearly show that
CO is detected significantly far away from the merging disks, 
and to possibly search for CO emission at the location of the 
tidal dwarf galaxy, several tens of kpc 
away from the merging disks. 
The typical integration time at each of these selected positions
is more than 3 hr. We have tried several positions 
at the tip of the southern tidal tail and
deep integrations in the possibly detected positions were further
conducted in 2000 March and April, totaling 27 hr usable
integration time in this tidal feature area. 

We used the dual SIS 3mmhi receivers connected with both the two 
256$\times$2 MHz filter banks and the two 600 MHz spectrometers
providing a velocity resolution of 2MHz $\sim 5$ \kms~ and a total
velocity coverage of 1330~\kms.
The system temperature $T_{\rm sys}$ (SSB) was typically less than
400~K (on a $T_R^*$ scale) and the weather 
conditions were excellent throughout almost all observing runs.
Occasionally, exceptional weather with 
$T_{\rm sys} \sim 200$~K (for a low-decl. southern sky source at 115 GHz!)
was seen in a few days. All observations were performed using 
a subreflector nutating at a chop rate of 1.25 Hz with a beam throw
of $\pm 3'$ plus a position switch (the so-called ``BSP'' mode)
to achieve the flat baselines. This ensures that the off-source
reference sky position is 6$'$ away from the observed on-source
pointing so that the telescope beam (FWHM$\sim 55''$) at the reference
position is well outside the CO extent of Arp~244.
Pointing and focus have been monitored
frequently every 1--2 hr by observing nearby quasars 3C279 or
3C273 (and occasionally planets at the beginning of the observations).
Uncertainties in positioning are typically $\sim 5''$. Calibration
with the standard chopper-wheel method was performed once
after every the other 
6 minute integration scan and yielded an antenna temperature on a $T_R^*$
scale. Further absolute flux calibrations in the antenna temperature scale
have been performed by the repeated observations of the northern nucleus 
(our map center) during different observing sessions and by the 
observations of some nearest well-observed starburst galaxies. 
Comparing the observed line strengths and profiles, we found that the 
consistency is satisfactory and the difference is less than $\sim 20\%$. 

HCN(1-0) was observed in 1997 April at only two locations: 
the northern nucleus (NGC~4038) and the CO peaks in the overlap 
region (Figure~1). This was part
of another project to search for the HCN emission in 
the LIG mergers. Additional 
integrations at these two locations
were obtained in 1999 November as a further consistency check. 
The 12m telescope's FWHM beam at 89 GHz ($\sim 72''$), pointed at 
the nucleus of NGC~4038, presumably covers all HCN emission in 
NGC~4038, yet excludes emission from that of NGC~4039 and most of the disk 
overlap region. The second HCN beam covers not only all the CO peaks and 
the extended CO in the overlap but also most CO emission
in NGC~4039. Therefore, 
the total HCN emission from Arp~244 can be roughly sampled by
simply summing up these two beam measurements. $T_{\rm sys}$ 
on a $T_R^*$~ scale
is about 230 K with 
the dual SIS 3mmlo receivers and the same back ends as used
in the CO observations. We have accumulated a total
integration time of nearly 5 and 7 hr for
NGC~4038 and the overlap/NGC~4039, respectively.

Our data reduction was performed using the CLASS/GILDAS package. 
Each individual
scan was checked for spikes/dips or bad channels and curvatures in
the baseline. Scans with the structured baselines have been abandoned, bad 
channels have been ``repaired'' by interpolation using the adjacent
channels, and a linear baseline was then subtracted after removal
of the spikes/dips for each accepted scan. All scans at each same 
position have then been summed to 
obtain an average spectrum. The data cubes of the different 
velocity spacings were also created from the highest velocity
resolution (2MHz $\sim 5$ \kms) and the smoothed (e.g., 
4MHz $\sim 10$ \kms) spectra maps. All spectra presented
in this paper have been smoothed to 8 MHz ($\sim 21~\kms$ for
CO and $\sim 27~\kms$ for HCN).

\section{RESULTS AND ANALYSIS}

\subsection{Previous CO(1-0) Observations}

The currently accepted total molecular gas mass in Arp~244 
is based on a single pointing observation with the 12m
in the overlap region. Sanders \& Mirabel (1985) listed
an integrated intensity of 15.9 K~\kms, or a total CO flux of only 
556.5 Jy~\kms ($\sim 35$~Jy/K on a $T_R^*$ scale), 
leading to a molecular gas mass of 2.6$\times 10^9$~\ms.
The Owens Valley Radio Observatory (OVRO) millimeter array's two 
overlapping field (FWHM$\sim 65''$) CO imaging only 
recovered $\sim 70\%$ CO flux (Stanford \etal 1990) of the
one-beam (FWHM$\sim 55''$) 12m measurement!
The 12m beam observed in Sanders \& Mirabel (1985)
only covered part of the overlap region, neither 
the nuclear regions of NGC~4038 and NGC~4039 nor the extended 
structures throughout the entire merging disks (see Figures~1 and 2).
Apparently, much of the CO is distributed over a much larger area
than the 55$''$ beam (Figure~2).
Furthermore, the integrated CO line intensity in the old 12m observation
was also underestimated owing to the insufficient velocity coverage (Figure~3,
more in \S3.2). 

Young \etal (1995) observed 6 positions in NGC~4038 and one position
in NGC~4039 with the Five College Radio Astronomy Observatory (FCRAO)
14m (FWHM $\sim 50''$)
and derived a total CO flux of $\sim 2070$ Jy \kms~
assuming an uniform CO disk of radius $\sim 0.'6$ for NGC~4038 
and an exponential CO disk of a scale-length 
$\sim 0.'32$ for NGC~4039. Aalto \etal (1995) observed 3 positions
in Arp~244 with the Swedish European Submillimeter Telescope (SEST)
15m (FWHM $\sim 45''$),
i.e., the overlap region and the nuclear regions of NGC~4038 and NGC~4039.
A sum of the integrated CO line intensities from the SEST observations
leads to a total CO flux close to what has been estimated by 
Young \etal (1995). 
Nevertheless, the new OVRO three-field imaging (Wilson \etal 2000) appears to
only recover a total CO flux of 910 Jy~\kms, but this is already
more than double that of the old OVRO map of Stanford \etal (1990).
All these recent measurements of still very limited
spatial coverage have already suggested that the total CO flux in Arp~244 can 
be a factor of a few larger than what has been previously accepted.

\subsection{New CO(1-0) Observations and the Total Molecular Gas Mass}

Figure~2 shows all the averaged CO spectra made with the 12m at a
half-beam spacing, a total of 73 spectra with more than 50 firm detections.
Only CO spectra at the outermost edge generally show non-detection or
tentative detections (noted in Table~1).
Many CO spectra in the inner part of the map (the disk 
overlap and the two nuclear regions) 
have the integrated line intensities more than $\sim 50 \%$
larger than that of the old 12m measurement of Sanders \& Mirabel (1985).
Observations in 1999 June at the same position as that of 
Sanders \& Mirabel (1985) gave an integrated line intensity 
exactly twice larger, consistent
with those mapped in the nearby positions in 1998 March/April. 
The northern nucleus was always
observed in all observing sessions giving consistent results
of about same integrated CO line intensity. 
Figure~3 presents the three spectra obtained 
near the CO peaks in the overlap region and compares with 
the original spectrum of Sanders \& Mirabel (1985).
Obviously, the old 12m observation 
may have suffered severely from the poorly determined baseline 
owing to the limited bandwidth, as the broadest CO spectra
in the overlap region have FWZI $\sim 600$~\kms. But, this is 
probably insufficient to explain a factor of 2 difference
in the integrated CO line intensity. Additional errors 
in the telescope pointing and calibration together with the limited
velocity coverage may have all contributed to the discrepancy.

The total CO flux, from our fully-sampled CO observations
(summarized in Table~1), is $S_{\rm CO} dV=3,172 \pm 192$~Jy~\kms~ 
($I_{\rm CO}=90.6 \pm 5.5$~K~\kms, using $ 35$~Jy/K conversion for 
the $T_R^*$~ antenna temperature scale), a factor of 5.7
larger than what was initially reported by Sanders \& Mirabel (1985).
The total CO luminosity is thus $L_{\rm CO}=0.3\times 10^{10}$~K~\kms~pc$^2$
($T_{\rm mb}$),
or a molecular gas mass (1.5$\pm 0.1$)$\times 10^{10}~\ms$
(using the standard CO-to-H$_2$ conversion factor of 
3.0$\times 10^{20}$~H$_2$~cm$^{-2}(\kkms)^{-1}$, or 
M(H$_2)=4.78\times (L_{\rm CO}$/K~\kms~pc$^2$)~\ms, applicable to 
GMCs in the Milky Way disk, \eg, Solomon \& Barrett 1991; 
Young \& Scoville 1991). The error estimate is purely from the
measurement uncertainties (statistical
errors from all measurements and the uncertainties in the CO detections
in the outer edge), additional uncertainties up to $\sim 20\%$
should be expected from the calibration and pointing etc. 

It is interesting to note that the new estimate of the dust 
mass in Arp~244, based upon the SCUBA 450 and 850 $\mu$m measurements
as well as the ISOPHOT observations at wavelength longer than 
100 $\mu$m, reveals a substantial
amount of cold and warm dust, and leads to a total dust
mass of $M_{\rm dust} \sim 10^8 \ms$ (Haas \etal 2000). 
Thus, for a total gas mass of $M_{\rm gas} \sim M({\rm H}_2)
+ M({\rm HI}) \sim 1.9\times 10^{10} \ms$ 
[$M({\rm HI}) \sim 4 \times 10^9 \ms$,
van der Hulst 1979; Hibbard \etal 2001 in preparation], a gas-to-dust ratio of
$M_{\rm gas}/M_{\rm dust} \sim 190$ is obtained. This is basically
the same gas-to-dust ratio established in our Galaxy
and in some nearby spiral galaxies like
NGC~891 (Alton \etal 2000). This simple consistency may also imply that
the standard CO-to-H$_2$ conversion determined from
GMCs in the Milky Way disk is likely applicable here and 
Arp~244 is indeed molecular gas-rich.
Table~2 summarizes the global quantities of Arp~244.

\subsection{Global Molecular Gas Distribution and Kinematics}

Although the most molecular gas is concentrated in the disk overlap
region, the extended distribution of the molecular gas 
spreads essentially all over the merging disks, much
further beyond the field of view of the WFPC2 (Figure~2). 
There are also concentrations of the molecular gas in the two
nuclear regions, particularly in the northern one.
This is better indicated in the quite coarse channel map of 100 \kms~ width
(Figures~4a--e). In general, there are good correspondences between the CO 
distribution and the optical morphology, except for the overlap region
where the CO emission appears to be much more prominent than the optical 
light. Figure~4f further compares the CO integrated intensity contours
(using all 74 position CO observations) with the optical disks.
All these (Figures~4a--4f) clearly show that the molecular gas extends
throughout the system and explain why the old 12m single beam measurement
underestimated the total CO emission by a large factor. Obviously,
much of the CO over a broad velocity range is simply distributed
over several arcminutes, corresponds to a linear CO extent of
possibly beyond $\sim $20~kpc.

The deep integration at the SE edge of the overlap region, 
where the southern tidal tail begins, confirms the clear detections 
of the weak CO emission from this 
region (Figure~5a). The central offset of the spectra shown in 
Figure~5a is (110$''$, $-82.5''$) relative to 
the northern nucleus, two full beams 
further out from the peak emission in the overlap region. CO emission
is also detected, for example, at offsets of (55$''$, 55$''$) and 
(81$''$, 27.5$''$) relative to the northern nucleus,
and in a few locations west of the two nuclear regions
at one and half beams from the nuclei (Figure~2 and
Table~1). These locations are so far away from
the CO emission peaks and their immediately associated CO extensions
(also see the new OVRO map of Wilson \etal (2000)), 
suggesting that there are weak extended CO features across over the 
entire optical disks. 
Given the relatively quite strong CO emission and the large
distance from the CO emission peaks, it is unlikely
that both the pointing errors, as large as 10$''$ in some rare
occasion, and the telescope 
beam pattern (sidelobes) can produce such relatively strong CO emission
that we have detected near the map edge. This certainly points out
the importance of the complete coverage of the merging systems in
the single beam observations since they may systematically 
underestimate the total CO content in the cases where the merging disks
are larger than the single beam. A substantial amount of the molecular
gas may be distributed at large galactic radii, especially in the young,
early and intermediate stage mergers like the ``Antennae'' galaxies.

The overall velocity spread of the molecular gas is more than 600 \kms~
across over the two merging disks. The narrowest velocity ranges are 
observed in the nuclear region of the northern face-on galaxy NGC~4038 
(FWZI~$\sim 300~\kms$, FWHM~$\sim 100$~\kms), whereas a broader 
velocity spread 
(FWZI~$\approxgt~500~\kms$, FWHM~$\sim 230$~\kms) can be seen in 
the southern inclined galaxy NGC~4039. 
The broadest velocity components (FWZI~$\sim$~600~\kms, FWHM~$\sim 300$~\kms)
are particularly prominent south of NGC~4039 even though the CO emission
is quite weak (Figure~2 and Nos. 6, 9, 17, 27, \& 29 in Table~1).
It is likely that the interferometric
observations have almost entirely resolved and thus missed these weak and
broad velocity features.
The velocity extent around the peak CO emission in the overlap
has a similarly broad velocity range.
Also, the velocity spread of most spectra in the 
overlap region lies between the two extremes observed in the two 
galaxies. This is obvious since some spectra contain the CO
contribution from either NGC~4038 or NGC~4039 or both. 
But, clearly most spectra at a full beam (or more) eastern of the nuclei
are basically free from any contribution of the nuclear CO emission.
We also notice that, although the difference is small ($\sim 50$ \kms), 
the systemic velocity in the overlap region is lower than that 
of the entire system whereas the systemic velocity
south of NGC~4039 is higher. This perhaps reminisces the disk 
rotation of the molecular gas in the more inclined galaxy NGC~4039.
Nonetheless, the average velocities of the molecular gas in 
the nuclear regions of both galaxies are about same as the systemic 
velocity of the entire system. 

The detailed study of the molecular gas distribution and kinematics
is difficult given the limited resolution of the 
12m telescope beam, which will be 
thoroughly described in the presentation of the BIMA + NRAO 12m 
full-synthesis observations, providing a much improved spatial
resolution of nearly 10 times better. Although the enormous amount 
of molecular gas in the overlap region must originate from 
both spiral galaxy progenitors, the kinematics and the
proximity of the molecular gas in the overlap, combined with 
the similar velocity spread of the molecular gas motion as that in
the southern galaxy NGC~4039, seem to suggest that
the most molecular gas in the overlap could be
originated from the southern progenitor. The molecular 
gas content of the two progenitors might be initially 
comparable and both were gas-rich. However, 
the VLA HI observations appear to show that the
progenitor of NGC~4039 was gas poor since the most HI in this system
is currently in the southern tail extending from 
NGC~4038 (Hibbard \etal 2001 in preparation). 

\subsubsection{CO in the Southern Tidal Dwarf Galaxy}

We have also searched for the faint possible CO emission at 
several positions in the southern tidal dwarf galaxy 
(Schweizer 1978; Mirabel \etal 1992) at the tip of the tidal tail. 
A possible weak CO detection from the beam at position
RA=12:01:26.0, Dec=$-$19:00:37.5 (J2000) appears to be at 
the $\sim 4 \sigma$ level (an integration of more than 12 hr; Fig.~5b).
An average of the summed spectra of all observations at the 5 
different positions we have searched (accumulated an integration
time of more than 27 hr), appears to confirm 
the weak CO detection at the $\sim 5 \sigma$ level. All these observed
positions are selected to be around the 
regions of the highest HI column density, where the VLA HI observations 
(Hibbard \etal 2000) have revealed some
large concentrations of the atomic gas,
distributed  extendedly around the region of the tidal dwarf
galaxy and along the southern tidal tail. 
The weak CO detections (Figure~5b) have also
shown the same velocity range as the atomic gas velocity spread,
revealed from the VLA HI channel maps, further indicating that the weak
CO emission is likely real. Yet the much deeper integration
and the more observing positions in this region are still 
required in order to firmly present the CO detections at the
very high significant levels, and to possibly obtain a better understanding 
of the weak CO distribution
and kinematics in a tidal dwarf galaxy.

The integrated CO line intensity from the detected position is
$I_{\rm CO}= 0.37 \pm 0.09$~K~\kms. The average spectrum of
the sum from all other 4 locations (not shown here) appears to give a similar
integrated CO line intensity $I_{\rm CO}= 0.35 \pm 0.12$~K~\kms,
while the average spectrum of all observations gives an integrated
CO line intensity $I_{\rm CO}= 0.35 \pm 0.07$~K~\kms.
Earlier observations by Smith \& Higdon (1994) failed to detect
any CO emission, but our observations here are much more sensitive.
Using the standard CO-to-H$_2$ conversion factor, we therefore
obtain a minimum molecular gas mass for the tidal dwarf
galaxy $\sim 0.61\times 10^8 \ms$ if only the detected position
is considered. It is likely that the weak CO
emission exists in other locations in this region and 
the total molecular gas mass for the entire tidal dwarf galaxy 
can be a factor of several larger. Indeed, the summed average spectrum
of all the 5 position observations seems to suggest a molecular gas mass 
$\approxgt~2\times10^8~\ms$. Anyway, this molecular gas mass
appears to be larger than that of the molecular complexes detected
in the M81 group (Brouillet, Henkel, \& Baudry 1992; Walter \& 
Heithausen 1999), yet comparable to the molecular
gas mass detected in the other two tidal dwarf galaxies (Braine et al. 2000).
It is possible that most of the molecular
gas might have come directly from the conversion of HI into H$_2$.
This is similar to the case in the two tidal dwarf galaxies 
where the molecular gas was detected in regions of the highest
HI concentrations (Braine et al. 2000), and is also related to the
case in the cold intragroup medium where huge HI concentrations and the
CO detection are both associated with the intragroup starburst 
(Gao \& Xu 2000) in the famous compact galaxy group Stephan's Quintet. 

\subsection{HCN(1-0) Detections and the Dense Molecular Gas}

The high dipole-moment molecules like HCN, which traces the dense molecular
gas at density $\approxgt~10^4$~cm$^{-3}$, are better tracers of the 
star-forming gas than that of CO~(Solomon \etal 1992; Gao 1996; 
Gao \& Solomon 2000a, 2000b). We have detected the weak HCN(1-0) emission 
from the two positions, one at the nucleus of NGC~4038 
and the other covering both the disk overlap and the nuclear
region of NGC~4039 (FWHM$\sim 72''$ at 89~GHz,
the two thick circles in Fig.~1). Figure~5c shows
the two HCN spectra and the total HCN luminosity
can be estimated by the sum of the two beam measurements.
This is because, in nearby galaxies where HCN maps exist, the 
dense molecular gas tends to be
concentrated in the innermost disks of the highest density 
regions (Gao 1996; Helfer \& Blitz 1997a). Thus it is unlikely
that any significant HCN emission in Arp~244 still exists outside 
the coverage of the two HCN beams. We estimated the total HCN flux to be
$S_{\rm HCN} dV=39 \pm 6$~Jy~\kms~($I_{\rm HCN}=1.2 \pm 0.2$~K~\kms,
using a $\sim 32$~Jy/K conversion) 
and the HCN luminosity $L_{\rm HCN}=0.7\times 10^8$~K~\kms~pc$^2$.
Therefore, the luminosity ratio of $L_{\rm HCN}/L_{\rm CO}=0.02$
implies only $\approxgt 5 \%$ of the molecular gas is at high density
of $\approxgt 10^4$~cm$^{-3}$ in Arp~244 (Gao 1996; Gao \& Solomon 2000a).
This is surprisingly low since ULIGs can have as high as 
$L_{\rm HCN}/L_{\rm CO}=0.25$ and up to about half of the molecular
gas at density $\approxgt 10^4$~cm$^{-3}$~(Solomon \etal 1992; Gao 1996; 
Gao \& Solomon 2000a). The ratio of $L_{\rm HCN}/L_{\rm CO}$ 
in Arp~244 is actually just comparable to that of the Milky Way disk 
(Helfer \& Blitz 1997b), even lower than that of
some ``normal'' spiral galaxies (Gao \& Solomon 2000a).
The fraction of the 
dense molecular gas in Arp~244 is apparently the lowest
among all LIGs observed so far. 

Although most of the CO emission ($\approxgt~ 60~\%$) is from the overlap
region, the HCN emission in the overlap appears to be less than half
since only about half of the HCN emission is originated from both NGC~4039
and the overlap. The HCN detection
from the overlap region and NGC~4039 has only a comparable integrated
line intensity, but weaker in the antenna temperature as compared with
the detection in NGC~4038. The two HCN spectra differ clearly in the
velocity spread: a narrow velocity space in the HCN emission of NGC~4038
whereas a much broader velocity spread in
NGC~4039 and the overlap region. Thus, the HCN spectra 
basically indicate that the dense molecular gas could have the 
similar gas kinematics as that of the total molecular gas revealed 
from the CO emission even though their spatial distributions are 
probably totally different. 

The low fraction of the dense molecular gas in Arp~244, especially
in the overlap region, is consistent
with the fact that the bulk of the molecular gas in the
overlap is extensively distributed both spatially and
kinematically (Figs.~2, 3 and 4). Although 
there are several CO peaks in the high resolution
synthesis maps, none of them has the extremely high 
concentrations of molecular gas as those found in ULIGs.
The highest gas surface density in the northern nucleus 
is $\sim 2.5\times 10^3$ \ms pc$^{-2}$ (resolved on a linear 
scale of $\sim 0.5$~kpc as in both the BIMA plus NRAO 12m
and the  new OVRO CO maps), order of magnitude
smaller than those in ULIGs, whereas
the gas surface density in the southern nuclear region is lower 
by at least a factor of 3 and comparable to 
that of the CO peaks in the
overlap. Our low-resolution (55$'' \sim 5$~kpc) 12m map  
reveals a CO peak in the disk overlap region since most of
the molecular gas is concentrated there. 
The peak molecular gas column density N(H$_2$) is close to 
10$^{22}$ cm$^{-2}$, i.e., an average gas surface density 
(over $\sim 5$~kpc scale) of about 190 \ms~pc$^{-2}$. This
low gas surface density together with the low HCN-to-CO line 
luminosity ratio clearly marks
the unique gas properties in Arp~244, in sharp contrast with those
in most LIGs/ULIGs.

\subsection{Local Star Formation Efficiency and the Radio-to-CO Ratio Map}

There is apparently a correlation between the radio continuum and the
CO emission as even so indicated in the old OVRO CO map (Hummel \& van 
der Hulst 1986; Stanford \etal 1990).
Particularly, most of the radio continuum emission is extensively 
concentrated in the overlap region with a very similar morphology 
to that of the CO. The direct comparison of the $KAO~ 60~\mu$m maps 
(Evans \etal 1997) with our full-synthesis CO images clearly reveals 
that the average FIR to CO ratio in the overlap region is higher 
than the global ratio since most of the FIR emission 
($\approxgt 75 \%$) is from the overlap region, 
whereas only $\approxgt 60 \%$ CO emission is from the same region. 
Therefore, the average SFE in the overlap can be a bit 
higher than the global value (which is
$L_{\rm IR}/M({\rm H}_2) =4.2~\ls/\ms$, or 5.3~\ls/\ms~ if a higher
100~$\mu$m KAO flux of Bushouse \etal (1998) is used), still comparable to
that of GMCs in the Galactic disk and an order of magnitude 
lower than that of ULIGs (Sanders \& Mirabel 1996; Solomon \etal 1997).
On the other hand, the average SFE in the nuclear regions is expected 
to be lower than the global value since
the two nuclear regions have extremely weak FIR emission, whereas the
CO is highly concentrated, particularly in NGC~4038.
Moreover, the radio continuum emission has indeed an excellent 
correspondence with the FIR maps, just as 
expected from the well-known correlation between the radio 
continuum and the FIR emission (\eg, Helou \etal 1985; Marsh \& Helou 1995).

In order to further quantify the local star formation properties 
and better identify the sites of the most intense starbursts with
the highest SFE, we have produced the ratio
map of the VLA 20cm radio continuum to the CO(1-0) emission. 
Since the tight FIR/radio correlation is also valid
on kpc scale in galaxies (\eg, Marsh \& Helou 1995; Lu \etal 1996),
the radio continuum emission can be used as a tracer of the
recent star formation in galaxies as well (\eg, Condon \etal 1990, 1991;
Condon 1992).
We here use the radio continuum maps to roughly indicate the
star formation sites. Thus, a high resolution indicator
of the FIR emission can be inferred by scaling 
the FIR according to the radio continuum emission. Comparing the
radio continuum maps with the detailed molecular gas distribution 
from the CO imaging, we can obtain the radio-to-CO ratio map to 
reveal the local SFE, which is defined as the
local ratio of the star formation rate to the molecular gas mass. 
Both the VLA 20cm continuum and the BIMA plus the NRAO 12m 
full-synthesis CO maps have about same resolution, thus a direct 
division can be performed to obtain the ratio map. 

The radio continuum to the CO ratios have been plotted
as contours and compared with the HST/WFPC2 (Fig.~6) and the
VLA radio continuum images (Fig.~7). Apparently, a rather 
smooth distribution of the radio-to-CO ratio
is observed across over most of the merging disks. 
The sites of the highest ratios, $\sim 2.5$ and 5 times above the average,
have been indicated in Figure~6 as black and white contours, respectively,
and they coincide well with the dusty
patches across over the HST/WFPC2 image. The sites of the highest 
radio-to-CO ratios are exclusively localized in the most prominent 
dusty regions in the overlap, which are near the CO peaks and 
probably the FIR emission peaks. The correspondent highest SFE 
is therefore $\approxgt 20~\ls/\ms$ since the global average SFE 
is 4.2 \ls/\ms.
Interestingly, almost all LIGs/ULIGs have SFE 
$\approxgt ~20~\ls/\ms$ (Sanders \& Mirabel 1996), although
some early stage pre-merging LIGs with SFE $\sim 10 ~\ls/\ms$
and lower do exist (\eg, Arp~302, Lo, Gao, \& Gruendl 1997).

Consistent with what can be roughly expected from a direct comparison
of the FIR map at 60~$\mu$m with the CO images in the nuclear regions, 
both nuclei have only a low radio-to-CO ratio, about the 
average or lower (Fig.~7). Thus the SFE in the nuclear regions 
is about 4 \ls/\ms. The $ISO$ MIR ``hotspots'' have higher ratios than 
the average, so do the extended overlap region and the western 
star-forming loop in NGC~4038, in addition to a few spots 
around the nuclear regions (especially the circumnuclear region of
NGC~4038, Fig.~6). 
But these are a factor of 2 lower than the highest
ratios and are only of moderate starburst 
sites with a SFE $\approxgt 10 ~\ls/\ms$, which is actually the
typical SFE value for the local starburst galaxies 
(Sanders \& Mirabel 1996).

\section{DISCUSSION}

Both observational and theoretical studies have demonstrated that
the ultraluminous extreme starburst phase is most likely achieved
in the molecular {\it gas-rich} mergers (e.g., Sanders \& Mirabel 1996;
Mihos \& Hernquist 1996; Downes \& Solomon 1998; Gao \& Solomon 1999)
where starbursts may proceed through the formation of numerous 
super-star clusters (e.g., Whitmore \etal 1999; Surace \etal 1998; 
Scoville \etal 2000). Multiwavelength observations of Arp~244 
and comparison with the early stage mergers observed by 
Gao \etal (1999) are therefore especially crucial to track how 
and when the dominant sources of star formation transform from 
within the disks of the two {\it gas-rich} spirals to the disk-disk
overlap regions between the two galaxies. Moreover, comparative study
with CO observations of the late stage mergers (Downes \&
Solomon 1998; Bryant \& Scoville 1999) are also equally important 
for understanding how the starbursts 
continue to evolve from taking place predominantly in the 
overlap regions to occurring mainly in nuclear regions 
when the merging advances to ULIG phase,
with extraordinary nuclear concentrations of dense gas. 
Intermediate mergers like Arp~244 serve apparently as an important
link in the merging process between a pair of gas-rich premergers
and the merged double-nucleus ULIG. We here discuss only
the starburst properties of Arp~244 related to our 12m observations.

\subsection{Ultraluminous Extreme Starburst Potential}

The CO content is decreasing as merging progresses, indicating the 
depletion of molecular gas due to merger-induced starburst 
(Gao \& Solomon 1999). Using the scaling of
M(H$_2) \sim S_{\rm sep}^{0.8}$ (Gao \& Solomon 1999, valid
for mergers at the early and intermediate stages, with the projected
nuclear separation 20~$\ga S_{\rm sep} \ga$~2~kpc),
the total molecular gas will decrease roughly by more than a factor of 2 
when Arp~244 reaches $S_{\rm sep}\sim$2 kpc,
prior to or about entering into the late merging stage. 
It appears that all ULIGs observed so far have 
$\sim 10^{10}$ \ms~ of molecular gas mass or more without exception, 
even after considering up to a factor of 5 reduced CO-to-H$_2$ 
conversion (Solomon \etal 1997).
The ultraluminous starburst phase of Arp~244 is thus possibly 
reachable in the advanced stage since only 
$\sim 10^{10}$ \ms~ molecular gas will be consumed by the newborn stars
in the next $\sim 10^8$ yr of the merging process. 
There will still be abundant
molecular gas of close to $\sim10^{10}$ \ms~ available, not to mention 
the additional amount of atomic gas in the merging disks and 
in the long tidal tails (Hibbard \etal 2001 in preparation) that may 
eventually fall back to the merging disks (Hibbard \etal 1994). 
The atomic gas could be an additional
gas reservoir for possible conversion into the molecular phase,
because there appears to be some evidence that ULIGs tend to have
the highest ratio of M(H$_2$)/M(HI) (\eg, Mirabel \& Sanders 1988, 1989).

In general, the gas surface density increases from orders of a few times
10$^2$ \ms pc$^{-2}$ to a few times 10$^3$ \ms pc$^{-2}$, as merging
progresses from early to intermediate stages~(Gao \etal 1999)
(unresolved on a scale of 1--2 kpc), whereas advanced ULIG mergers
such as Arp~220 have gas surface density typically greater than 
$10^4$ \ms pc$^{-2}$~(Scoville \etal 1997; Downes \& Solomon 1998;
Sakamoto \etal 1999). 
Although the total molecular gas content of Arp~244 is comparable
to that of ULIGs, the gas surface density is still orders of 
magnitude lower than that of ULIGs, e.g., in the overlap, 
$\sim 10^2$~\ms~pc$^{-2}$ on a scale of 5 kpc and 
$\sim 10^3$~\ms~pc$^{-2}$ on a resolved scale of sub-kpc.
Therefore, only a small fraction of the molecular gas in Arp~244
is actually at a high enough gas surface density for a high SFE, 
as formulated in the Schmidt law (\eg, Kennicutt 1998). 
The overall gas density is low to maintain a low SFE across
over the entire system, except for some localized starburst regions (Fig.~6),
just as our HCN observations have revealed that little dense molecular
gas available currently to power the extreme starbursts. 

Ultraluminous extreme starbursts require not only 
a large quantity of molecular gas but 
also a high gas concentration, particularly the nuclear
gas concentration, so that the bulk of the 
molecular gas is at a high density. The large quantity of pre-existing 
molecular ISM is the sufficient fuel for Arp~244 to ultimately 
reach the onset of the ultraluminous starburst phase, as the
merging proceeds to the advanced stage,
and the bulk of the molecular gas becomes highly condensed.
Unlike most LIGs/ULIGs, which show the dominant nuclear MIR and
perhaps FIR emission (Hwang \etal 1999; Soifer \etal 2000; 
Bushouse \etal 1998; Xu \etal 2000), Arp~244 has only 
moderate nuclear emission in MIR (Mirabel \etal 1998;
Xu \etal 2000), and the $KAO~60~\mu$m emission from the two nuclei 
is extremely weak (Evans \etal 1997). 
Although there is no clear indication of a large stellar bulge in either
of the galaxies, the concentrations of the molecular gas 
as well as the stars in the nuclear regions in Arp~244 perhaps
also draw some analogies to the case of galaxy-galaxy merger model 
with stellar bulges in the progenitors (Mihos \& Hernquist 1996). 
In reality, Arp~244 is probably more in between the two extreme cases
modeled by Mihos \& Hernquist (1996). According to 
the models (see Figure~5 in Mihos \& Hernquist 1996),
the first peak of starburst phase has recently passed in Arp~244, 
shortly after the pair's first closest approach, 
perhaps as evidenced by numerous young star clusters in the HST/WFPC2 images. 
Thus, Arp~244 could have just phased out the bona fide starburst
LIG stage. As more and more gas continues
to accumulate in both the overlap and the two nuclear regions, 
the further nuclear gas concentrations and the continuous star formation 
can build stronger bulges that
can help stabilize the disks against the bar formation, lower the 
SFE levels during the merging, and leave enough gas for 
the strongest final starburst when the two galaxies eventually 
coalesce, with almost all the molecular gas collapsed into 
a kpc double-nucleus region. More than an 
order of magnitude increase in the SFE, combined with an 
abundant molecular gas supply of $\sim 10^{10}$~\ms,
will be just sufficient to make Arp~244 into $L_{\rm IR}\sim 10^{12}$~\ls.
Interestingly, there also appears to be some quite widely separated
ULIGs ($S_{\rm sep}>$~10~kpc), which might still be in their
early stage of merging, but may have passed their first closest 
impacts, and may have just been experiencing their first strongest starbursts 
(Murphy 2000).

In summary, a preexisting
abundant molecular gas content is a necessary for there to be 
a possibility of a merging spiral pair reaching
the ultraluminous starburst phase. Arp~244, with a large quantity 
of molecular gas available at low density, a {\it low} SFE, a moderate
FIR luminosity or star formation rate, and yet a relatively
early/intermediate stage of merging, has 
the potential of producing an ultraluminous extreme starburst 
in late stage of merging,
when the molecular gas complexes in the overlap region
eventually merge with the nuclear gas concentrations
and finally collapse into a kpc-scale double-nucleus gas disk. 

\subsection{Starbursts and the Current Intense Star Formation Sites}

The well-known correlation between the
FIR thermal dust emission and the radio continuum 
synchrotron emission, unexpected as they are apparently 
two unrelated physical mechanisms, is probably the tightest 
relation known among the global quantities of galaxies (e.g.,
Helou \etal 1985; Condon 1992; Xu \etal 1994). The morphologies of 
the two emission are also similar and the correlation 
between them appears to hold down to kpc scale within individual galaxies 
(e.g., Bicay, Condon, \& Helou 1989; Marsh \& Helou 1995; Lu \etal 1996).
Based on these, we approximately scaled the local 
FIR emission according to the radio continuum emission, and compared
with the molecular gas distribution to measure 
the local SFE. We have clearly identified that the most intense bursts
of star formation (the sites of the highest SFE, i.e., the highest
radio-to-CO ratios, Figure~6), currently being observed, are not in the 
concentrations of the optically revealed star clusters, or even
the peaks of the MIR emission,
nor necessarily at the exact locations of the CO and 20cm radio 
continuum emission peaks. Instead, the strongest starbursts 
appear to be confined in small regions near the CO, radio continuum, 
and FIR emission peaks in the overlap region.
The bulk of the molecular gas across over the entire system is 
making stars with a rather modest or ``normal'' SFE (Fig.~7). 

Let us emphasize what we have learned from our observations and 
the various other observations.
Starbursts have apparently been going on or happened
in some regions in Arp~244 
as evidenced by super-star clusters~(Whitmore \etal 1999) and 
the MIR ``hotspots'' in the overlap~(Mirabel \etal 1998).
Yet, most of the FIR emission (as well as the MIR), 
which dominates the total energy
output, comes from the overlap region (Evans \etal 1997;
Bushouse \etal 1998), where most molecular gas resides, rather than from 
the super-star cluster concentrations.
Although Arp~244 has long been claimed as the 
nearest archetypal starburst merger, our observations
show that the molecular gas is rather extensively distributed over
the entire merging disks, though most is still in the overlap, and 
the bulk of the molecular gas is only making stars at a low
SFE. The entire system is presently not undergoing 
a global burst of star formation, even though some localized starbursts
are occurring in the overlap, and a global starburst 
could have just peaked recently, right after the first closest
impact responsible for the formation of the tidal tails. 

Whitmore \etal (1999) found that the star clusters at the edge of 
the dusty overlap region appear to be the youngest, with ages 
$\approxlt 5$ Myr, while the star clusters in the western loop in NGC~4038
appear to be 5-10 Myr old. Indeed, these are the sites of the 
most intense and moderate starbursts respectively as revealed 
by the radio-to-CO ratio
map in Figure~6. There might be many youngest star clusters
totally obscured by the dust at the most intense starburst sites, 
contributing significantly to
the dominant energy output in the overlap region. On the
other hand, many star clusters in the northeastern 
that appear to be $\sim 100$~Myr old, and even older
star clusters across over the system, are not the current starburst
sites, contributing little to the total FIR emission. 
These old star clusters are probably related to the recent large-scale 
active star formation happened after the first closest encounter. 

The excellent agreement among the
sites of the highest radio-to-CO ratios, the dark patches 
in the HST/WFPC2 images, and the youngest ages of the 
star clusters, is striking. 
Apparently both the most intense localized starbursts and the
overall large-scale star formation are occurring in the overlap
region, where all the CO, [C~$_{\rm II}$] line (Nikola \etal 1998), 
FIR, and radio continuum show high concentrations. But, the 
confined intense starburst sites, which are heavily obscured in 
the dust, have only SFE $\approxgt 20~\ls/\ms$, just reaching
the typical SFE level for LIGs/ULIGs. On the other hand, the average 
SFE in the overlap is just slightly larger than 4~\ls/\ms,
still comparable with that of the Galactic disk GMCs.

One caveat is that Arp~244 may be twice luminous in 100 $\mu$m,
as measured by the KAO (compared with
that of the $IRAS$ measurement, Bushouse \etal 1998),
although the 60 $\mu$m measurements agree (Evans \etal 1997).
This discrepancy needs future FIR observations to resolve.
More importantly, it is perfectly
possible that the standard CO-to-H$_2$ conversion factor can be 
an overestimate for the total molecular gas mass. Therefore, it is
likely that the global SFE in Arp~244 can be up to more than
a factor of 2 larger than that of the GMCs in the Milky Way disk.
This may indeed suggest that an enhanced global 
SFE proceeds throughout
the entire merging disks while some confined
strongest starbursts have a highest SFE much larger than 20~\ls/\ms,
typical for ULIGs. In any case, the highest
radio-to-CO ratio sites could actually be bona fide
sites of the {\it current} starbursts with an elevated SFE, and
may contribute a significant fraction to the
total FIR emission, even though most of the 
FIR emission may still come from the regions of
a normal or modest SFE.

Using the [C~$_{\rm II}$]/CO(1-0) line ratio to 
distinguish between starburst
activity in galaxies and more quiescent regions (Stacey \etal 1991),
Nikola \etal (1998) concluded that there is no strong starburst
activity taking place in Arp~244 on a scale of $\sim 5$~kpc,
the resolution of their data. This is roughly in agreement 
with our results since starburst sites are only identified on
a scale of $\approxlt 1$~kpc (Fig.~6). Nikola \etal (1998) 
further argued that most of the [C~$_{\rm II}$] emission might have
come from the confined active star forming regions surrounded with 
the more quiescent GMCs.
Our results appear to support this claim if indeed the 
sites of the intense starbursts, which are identified from 
the high radio-to-CO ratios (Fig.~6), are producing the bulk of 
the [C~$_{\rm II}$] emission. Additional supports for the
confined starbursts in small regions in the overlap
are the recent ISO-SWS measurements (Kunze \etal 1996) and the ISO-LWS
observations, including the detection of Br{$\gamma$} knots in
the overlap interaction zone (Fischer \etal 1996).

The rather extended moderate starbursts also coexist closely
with the confined, most intense starbursts
in the overlap region. Other interesting sites of moderate starbursts
are the western loop/ring structure, and part of the $\approxgt$~kpc scale
circumnuclear regions (mainly in NGC~4038), but not the nuclei themselves
(Figs.~6 \& 7), with nearly 3 times higher SFE than 
the average. The western loop coincides with the molecular ring
we have mapped from the BIMA plus the NRAO 12m full-synthesis CO image, 
which was completely absent from the old OVRO CO map (Stanford \etal 1990),
and was only partially revealed in the new OVRO CO map (Wilson \etal
2000). It appears that the starbursts occur at a moderate level in the
southeast side of the overlap region, proceed progressively with 
increased intensity 
towards northwest across over the overlapping molecular concentrations, 
and produce the most vigorous starburst
in the northern and western edges of the huge molecular gas 
agglomerations in the overlap interaction zone (Fig.~7).

Moderate [C~$_{\rm II}$] emission is also detected in the western loop 
(Nikola \etal 1998), which also
indicates the moderate star-forming activity there. This is again consistent
with our results obtained from the radio-to-CO ratios. Moreover, 
the [C~$_{\rm II}$] 
emission peak appears to be slightly offset from our single-dish CO 
peak (Figures~2 \& 4f), which is resolved into several
CO peaks by the interferometers. The strongest [C~$_{\rm II}$] 
emission seems to be roughly coincident with the sites of the highest SFE, 
just north of the CO peak in the NRAO 12m map. Our single-dish CO map 
also appears to be
roughly peaked at same position as that of the KAO 60 (Evans \etal 1997),
100 and 160 $\mu$m
maps (Bushouse \etal 1998). The strongest MIR peak in the
overlap region that is inconspicuous at optical wavelength (Mirabel
\etal 1998; Xu \etal 2000), however, may correspond only to the southernmost
CO peak in the overlap (Wilson \etal 2000), which is not one of
the strongest starburst sites and has only a moderate
SFE~$\approxgt 10~\ls/\ms$ (Figures 6 \& 7). Therefore, 
the strongest peaks of the CO, radio continuum,
and FIR, as well as the MIR peaks, are not necessarily
the exact sites of the highest SFE, or the most intense starbursts. 
The highest radio-to-CO ratio peaks, which may
reveal the sites of the most vigorous star formation with the highest SFE,
and the strongest [C~$_{\rm II}$] emission peaks, which arise mainly from the
photodissociation regions, appear to be the best probes of 
the most intense starbursts, which might be totally obscured in 
the optical/near-IR and even in the MIR regime.

\subsection{Star Formation from the Multiwavelength Observations}

In light of the numerical models for mergers of Arp~244 alike (Toomre \&
Toomre 1972; Barnes 1988; Mihos \etal 1993; Mihos \& Hernquist 1996;
Barnes \& Hernquist 1996, 1998) and the multi-wavelength
observations of Arp~244, comparison of the various observations 
and statistical studies of LIG mergers  with those models can be fruitful.
We here try to discuss qualitatively the onset of 
the starbursts, the fate of the starbursts, and the possible 
ultraluminous extreme starburst phase in Arp~244,
utilizing these observational and theoretical results.

ROSAT high-resolution imager (HRI) X-ray map 
(Fabbiano \etal 1997) and the high-resolution and high-sensitivity 
Chandra X-ray images from the Chandra news release (Fabbiano et al. 2000),
after the submission of this paper,
show some correspondence with the discrete 
radio knots. Most soft X-ray emission, however, is from the disks 
and nuclear regions, rather than from the overlap region, even though 
several X-ray emission knots exist in the overlap. 
Yet, more soft X-ray structures in regions of the $ISO$ ``hotspots''
are evidently better revealed in the new Chandra X-ray images. Both the 
X-ray and radio continuum show extended morphology (prominent ring 
structure) in the northern galaxy NGC~4038, though not in
NGC~4039. Since
supernova remnants are most likely responsible for the majority 
of both extended emission, these observations may be used as
probes to identify {\it past } active 
star formation sites. According to the various models, 
the first burst of star formation may have occurred
several $\sim 10^8$ years ago when the first
closest approach of the galaxy pair happened. Many X-ray knots 
with the radio emission correspondence can be even older, and they are
reminiscent of the possible star forming sites in the past, prior to or
during the first strongest impact. Overall, these emission could mark roughly
the sites of the interaction enhanced large scale star formation
across over the pair's disks at the earliest stage of merging.

H$\alpha$ emission and young star clusters shown in the HST images 
(Whitmore \etal 1999) indicate
that recent vigorous star formation is also proceeding throughout 
the entire merging disks. Molecular gas concentrations in
the overlap may have hidden some extremely
young star clusters and HII regions. Nevertheless, 
these again are the rather {\it recent} large scale bursts of star 
formation occurring probably right after 
the genesis of the tidal tails, 
after the pair's first closest approach.

Both the MIR and FIR observations (Mirabel \etal 1998; Xu \etal 2000;
Evans \etal 1997; Bushouse \etal 1998) show the strongest peaks in
the rather extended overlap region where the bulk of the molecular gas resides.
But the MIR and FIR emission peaks differ from each other, whereas
the FIR peaks roughly at same position as that of the CO.
Generally speaking, the overlap region is the {\it current} sites 
of vigorous star formation or starburst.
Yet, the overall starburst  level is at most modest. The strongest
starburst sites with SFE comparable
to that of LIGs/ULIGs are confined to $\approxlt$~kpc scale small regions 
in the overlap, and the global SFE level is about same as those 
in the Milky Way disk GMCs. Despite some localized starbursts are
still going-on, the recent peak of starburst phase in Arp~244 has most likely
passed already, and the entire system is currently in a rather
quiescent star formation phase.

The radio continuum-to-CO ratio map implies low SFE at both nuclei 
(Fig.~7). The highest gas concentration in Arp~244
revealed by the interferometers is, however, in the northern 
nuclear region, which
has only moderate $\approxgt$~kpc scale circumnuclear starburst ring,
rather than a sub-kpc nuclear starburst. The highest radio-to-CO
ratios are the confined bona fide starburst sites (Fig.~6) which
appear to have some correspondence with the [C~$_{\rm II}$] emission
peaks (Nikola \etal 1998).
The CO emission peaks in the overlap region appear not to be
at the exact locations of the localized starbursts, and the most 
intense starbursts might not be at an extremely high SFE level 
as in ULIGs. These highest molecular
gas concentrations are likely to undergo
{\it future} ultraluminous extreme starburst once they all merge 
together with the nuclear gas concentrations,
and collapse into $\sim $~kpc scale in the final coalescence. 
Although the progenitor galaxies of Arp~244
could only have small stellar bulges, the bulges can be built up 
through merging (\eg, Carlberg 1999) in addition to
nuclear gas infalls and concentrations.
Thus, Arp~244 fits most likely between the two extremes
modeled by Mihos \& Hernquist (1996), and more than an order 
of magnitude enhanced SFE can be achieved in the final merging.

Although most ($\ga 60\%$) HI gas is distributed 
in the tidal tails~(van der Hulst 1979; Hibbard \etal 2000), 
there are still $\sim 2\times 10^9 \ms$ HI
in the merging disks, whereas most ULIGs have little HI left in the 
merged disks~(Hibbard \& Yun 1996, 2000; Mirabel \& Sanders 1988, 1989).
Most HI in Arp~244 is in the southern tail, where we tentatively 
detected CO emission at the tip of the tidal tail, indicating 
a molecular gas mass of $\approxgt 2\times 10^8$~\ms~ in the 
tidal dwarf galaxy.
These are additional gas reservoir for {\it future} star formation, 
especially when the extraordinary HI tails rain back onto the 
merging disks (Hibbard \etal 1994), converting most HI gas
into molecular phase.

Our recent HI observations of NGC~6670 suggest that a precursor
to Arp~244 can be that the HI disks have merged into a huge overlap
concentration in the extended HI disks, prior to the
merging of the stellar and molecular gas disks~(Wang \etal 2000).
In Arp~244, although an anti-correlation 
between the molecular and atomic gas distributions appears
to exist, there is still a significant amount of HI gas in 
the molecular gas overlap region. Whether there was a HI gas overlap
formed previously, prior to the merging of the molecular gas disks, 
requires high resolution HI observations and further comparative studies 
of the multi-wavelength observations, in combination with the numerical 
modelings, to test. Clearly, understanding of the formation and 
evolution of the overlap region is the key to ultimately
comprehend the entire star formation history of Arp~244.

\subsection{Overlap Star Formation and Starburst Mechanisms}

The formation of the overlap star formation regions, unnecessarily as 
overwhelmingly dominant as in the case of Arp~244, might be a quite 
common phenomenon in merging galaxies (Xu \etal 2000). Both the 
simulations and observations have 
shown that the gas is being transported into the
nuclear regions when a pair of spiral galaxies undergoes through
the merging process, in addition to being dragged out into the tidal
tails (\eg, Olson \& Kwan 1990a,b; 
Noguchi 1991; Mihos \& Hernquist 1996; Barnes \& Hernquist 1991, 1996;
Scoville \etal 1991; Gao \etal 1999; Hibbard \& van Gorkom 1996). Yet, models 
appear to be unable to reveal the early formation of the gas concentration
in the overlap region when galaxies are still quite widely separated,
like the intermediate merger Arp~244 with the molecular gas overlap, 
where most gas resides, and the early merger NGC~6670 with the atomic 
gas overlap. Could a significant amount of the atomic gas, especially 
in the southern progenitor, have ended up in the overlap region in 
the first place, prior to the formation of
the molecular gas overlap, as a result of efficient 
HI cloud-cloud collisions than the collisions of the GMCs, 
when the two gas-rich spiral 
progenitors first collided? This is because the mean free path
of a HI cloud is $\sim 33$~pc, much smaller than the size of the
overlap region, yet the GMCs' mean free path is orders of magnitude
larger and the chance of GMC collisions in the overlap is 
extremely small (Jog \& Solomon 1992). 
Just like the early merger NGC~6670, which has the HI gas 
concentration in the HI disk overlap region formed prior to 
the merging of the stellar and molecular gas disks (Wang et al. 2000),
there might be a similar HI gas overlap formed in Arp~244 during
its earliest merging stage.

Jog \& Solomon (1992) argued that a starburst 
occurs when  the preexisting GMCs in the overlap 
region undergo radiative shock compressions by the preexisting 
high pressure of the central molecular intercloud medium 
produced by the heating from the HI cloud-cloud collisions. If indeed
an HI disk-disk overlap region could be formed prior to 
the merging of the stellar and molecular gas disks,
the dominant star formation should not be occurring in the HI gas
overlap region, given few GMCs exist in this HI cloud-cloud 
collision ISM during the early stage of merging. Instead, large-scale
star formation and nuclear starbursts induced by the gas inflow 
are probably at play (Combes \etal 1994).
It is later formation of the molecular gas disk overlap, within
this preexisting high pressure overlapping HI gas concentration, 
when the merging advances into the intermediate stage, that will
probably make GMCs in the overlap undergo starburst by 
the radiative shock compressions (Jog \& Solomon 1992). 

Although the soft X-ray emission in the overlap region in Arp~244
is neither strong nor extended, as revealed by
the ROSAT HRI (Fabbiano \etal 1997), this may be due to 
the high gas column density in the overlap 
which absorbs the most extended soft X-ray emission.
The Chandra soft X-ray images of much improved sensitivity
and resolution do show some extended features in the
overlap, which are particularly in good correspondence with 
the MIR ``hotspots''. This is in the right track, just as expected
from the prediction of the overlap starburst model of Jog \& Solomon (1992).
Higher energy X-ray imaging that can penetrate through the 
heavy dust/gas concentration in the overlap, yet still orders of 
magnitude below the Compton-thick limit, and the extinction-corrected
soft X-ray images of Arp~244 could be the key 
to further test the overlap overpressure starburst mechanism.

In advance mergers, the double-nucleus gas disks are already merging
with the overlap gas concentrations, with most gas either between
(\eg, NGC~6240, Tacconi \etal 1999) or around the double-nucleus
(\eg, Arp~220, Downes \& Solomon; Sakamoto \etal 1999) in
$\sim$kpc scale, in addition to the nuclear gas concentrations. 
Thus, a possible merging sequence for a gas-rich pair of spirals 
may look like the following: from the build-up of the HI gas disk 
overlap regions (NGC~6670) $\to$ the sufficient overlap in both 
stellar and molecular disks (Arp~244) $\to$ the kpc scale merging
double-nucleus gas disks, either with a central gas concentration
between the nuclei (NGC~6240), or with an extended gas disk of 
several kpc surrounding the nuclei (Arp~220). The transformation
of the merging phases and the further development of the overlap 
regions are also accompanied by the change in the dominant sources 
of the total power output: from large scale star formation within 
galaxy disks in early mergers $\to$ the mainly localized 
starbursting overlap regions between the galaxies in intermediate 
mergers $\to$ the extreme starburst
in  the highly concentrated double-nucleus sub-kpc sources in 
advanced ULIGs.

The overpressure starburst mechanism (Jog \& Solomon 1992)
is probably happening in intermediate mergers, especially those 
with the extensive overlap regions like Arp~244. Yet, this 
mechanism appears to require a sufficient overlap between the gas disks 
so that the early formation of the gas overlap region is possible.
Is this the sole starburst mechanism during the entire merging
process when two gas-rich spiral galaxies merge? 
It is likely that different starburst mechanisms play 
various roles during different merging stages along a merger 
sequence. What then are the other triggering mechanisms for the starbursts?
Wilson \etal (2000) compared their new OVRO CO map with the
$ISO$ MIR emission and suggested that molecular cloud collisions may play
an important role for the local intense starburst and the strong 
MIR emission. Yet, Lo \etal
(2000) found no obvious signatures of cloud collisions
from analyzing the molecular gas kinematics in the BIMA + the NRAO 12m
full-synthesis CO data.
Perhaps Arp~244 is on its way of changing from the overpressure overlap
starburst to the direct cloud-cloud collision starburst. 

An extensive active star forming molecular gas overlap region
observed in Arp~244 may not always be the case
for most mergers. In fact, most LIGs do not show such an extensive
molecular gas overlap region 
(\eg, Gao \etal 1997, 1999; Bryant \& Scoville 1999). Even though
star formation in the overlap can be common in
mergers, the overlap starburst is probably much 
less dramatic than that of Arp~244 in most cases (\eg, Xu \etal 2000).
But, do all gas-rich mergers develop such an extensive
molecular gas overlap phase during the course of merging? We do not have
a definite answer. Only in late merging stage have both observations
and numerical simulations demonstrated the existence of the
highly concentrated molecular gas between the merging nuclei.
Also, only in advanced mergers, can the chance 
of GMC-GMC collisions be much enhanced since almost all 
the molecular gas (in both the overlap and the two nuclear gas concentrations)
finally condenses into a kiloparsec scale.
Now, the mean free path of GMCs is less than
the size scale of the gas concentration region since the volume
filling factor of GMCs is ten-fold increased, making even 
the intercloud medium molecular. Therefore, direct collisions of GMCs could 
be playing an important role for the enhanced starburst 
when merging progresses into advanced stage in the final coalescence. 

\section{CONCLUSIONS}

We summarize our main results and present our 
concluding remarks in the following:

1. Our fully sampled CO(1-0) map (FWHM$\sim 55''$)
at half-beam spacing reveals a factor
of 5.7 larger total CO flux in Arp~244 than the 
previously accepted. This is mainly due to the enormous CO extent
of Arp~244, which is much larger than the single pointing $55''$ 
beam used in the old observation. In general, 
single beam observations of merging
systems may well systematically underestimate CO content in the cases 
where the merging systems are much larger than the single beam, unless
a map of extensive coverage of the merging systems is made.
Most of CO emission in Arp~244 indeed peaks at the disk overlapping 
region with a broad velocity spread of 600~\kms.
We obtain a total molecular gas mass of 1.5$\times 10^{10}\ms$
for Arp~244 using a standard CO-to-H$_2$ conversion factor.

2. HCN(1-0) emission has been detected in Arp~244. HCN observations
suggest that there is only a small amount of dense molecular gas
in Arp~244 compared to the total molecular gas content. The fraction
of the dense molecular gas is only comparable to that of 
normal spiral galaxies and is much lower than that of
luminous and ultraluminous infrared galaxies (LIGs/ULIGs).

3. We detected extended CO emission far away (nearly two full
beams) from the major CO concentrations, such as those revealed by the 
interferometers. The peak CO emission in the overlap region has an 
average gas surface density near $\sim 200~\ms$~pc$^{-2}$ 
over $\sim 5$~kpc --- the scale probed by the $55''$ beam.
Even the resolved interferometry CO images only 
reveal that the highest gas 
surface density is about 10$^3~\ms$~pc$^{-2}$, still
order of magnitude lower than that of ULIGs.
This is consistent with the small fraction of dense gas found
from the HCN measurements. Weak CO emission from the
tidal dwarf galaxy at the tip of the southern tail has  
also been tentatively detected and the estimated molecular
gas mass is possibly larger than $\sim 2\times 10^8$~\ms.

4. The excellent correspondence of the radio continuum images 
with the FIR maps is evident in Arp~244, which also
implies the tight correlation between the two emission.
We use the high resolution VLA 20cm image to approximately
represent the FIR emission distribution in Arp~244, and compare
with the CO images of comparable resolution to estimate
the local star formation efficiency (SFE), which is defined as 
the FIR to molecular gas mass ratio, and is now indicated by 
the radio-to-CO ratio. SFE is low throughout 
the entire system with a global SFE$=4.2~\ls/\ms$, just 
same as that of the Galactic disk GMCs.
Only some confined regions in the overlap, inconspicuous
in optical observations,
are the sites of the current powerful starbursts with
SFE~$\approxgt 20~\ls/\ms$, comparable to that of LIGs/ULIGs. The VLA
radio continuum images also reveal a very similar 
morphology as that of the molecular gas distribution. 

5. Although some localized starbursts are going on in the confined 
regions in the overlap, the bulk of the molecular gas is forming 
into stars with a low SFE, just same as that of normal spiral 
galaxies. The large amount of 
total molecular gas, yet only a small amount of dense molecular gas 
and a low SFE, revealed from our observations, combined with the 
theoretical predictions of models of gas-rich mergers, 
seem to suggest that Arp~244 has
the ultraluminous extreme starburst potential when merging
proceeds into late stage. Arp~244 appears to be a snapshot of 
an ULIG in its early stage of making.

6. Multi-wavelength comparison together with numerical models appear
to indicate that Arp~244 has recently passed the peak of large
scale active star formation. Globally, the bulk of the molecular gas
in Arp~244 is currently not in a starburst phase. The confined
bona fide starburst sites of the highest SFE appear to be offsets from 
the peaks of the CO, radio continuum, FIR, and MIR emission. 
The future ultraluminous
extreme starburst phase can possibly be reached once the CO 
concentrations in the overlap merge with the double-nucleus 
gas concentrations in the final coalescence.

7. In conclusion, our newly determined 5.7 times larger total 
molecular gas mass in 
the ``Antennae'' galaxies lowers the global SFE
to be comparable with that of GMCs in the Milky Way
disk. Comparing the CO images with other observations, 
including our own VLA 20 cm continuum imaging,
an overall relatively low molecular gas density
and a low SFE across over the bulk of the molecular gas distribution
have been revealed, which is in sharp contrast 
with most LIGs/ULIGs. The strongest starburst sites are currently
confined in small regions heavily obscured in the dusty patches.
We conclude that the large reservoir of 
molecular gas is the sufficient fuel for Arp~244
to ultimately enter an ultraluminous extreme starburst phase
as merging proceeds to the advanced stage. The ultraluminous phase 
is possibly achieved once almost all molecular gas condenses into 
$\sim$kpc scale, with an order of magnitude increase in the gas 
surface density and more than ten-fold higher SFE.

\acknowledgements

We thank the NRAO 12m staff for generous supports and additional
allocations of the observing time. We are grateful to the referee
David Sanders and the Scientific Editor Gregory Bothun for helpful
comments and suggestions. We also thank Robert Gruendl for assistance 
with some of the observations and helpful suggestions to an earlier 
version of this paper.
Y.G. \& K.Y.L. acknowledge support from 
the Laboratory of Astronomical Imaging which is funded by NSF grant
AST 96-13999 and by the University of Illinois. 
K.Y.L., T.H.L., \& S.W.L. acknowledge support from the Academia Sinica.
Y.G. is also grateful to E.R. Seaquist for various support 
at the University of Toronto.
This research has made use of the NASA/IPAC Extragalactic Database (NED) 
which is operated by the Jet Propulsion Laboratory, California Institute of 
Technology, under contract with the National Aeronautics and
Space Administration. 
Y.G. is currently supported by the Jet Propulsion Laboratory,
California Institute of Technology, under contract with NASA.

\clearpage

\begin{deluxetable}{rrrrrr}
\tablecaption{Summary of the NRAO 12m CO Mapping Results \label{tbl-1}}
\tablehead{
\colhead{No.}         &   
\colhead{R.A.\tablenotemark{a}}  &  \colhead{Decl.\tablenotemark{a}} &
\colhead{$I_{\rm CO}$\tablenotemark{b}}  &   \colhead{$V_{\rm CO}$} &
\colhead{$\Delta V_{\rm FWHM}$} \\
\colhead{ }               &
\colhead{arcsec} & \colhead{arcsec} &
\colhead{K~\kms}  &   \colhead{\kms} &
\colhead{\kms}
}

\startdata

 1 &   0.0 &   0.0 &  19.5 & 1632 &  93 \\
 2 &   0.0 &  27.5 &  11.9 & 1612 & 204 \\
 3 &   0.0 & $-$27.5 &  20.1 & 1607 & 201 \\    
 4 &   0.0 &  55.0 &   5.2 & 1607 & 282 \\    
 5 &   0.0 &  82.5 &   3.5 & 1561 & 331 \\    
 6 &   0.0 & $-$82.5 &  10.6 & 1659 & 281 \\    
 7 &   0.0 & $-$55.0 &  19.2 & 1604 & 227 \\    
 8 &   0.0 & 110.0 &   0.4\tablenotemark{c} & 1663 & 110 \\    
 9 &   0.0 &$-$110.0 &   3.9 & 1638 & 348 \\   
10 &   0.0 &$-$137.5 &   1.3 & 1660 & 229 \\    
11 &  27.0 &   0.0 &  21.5 & 1593 & 198 \\    
12 &  27.0 &  27.5 &  13.6 & 1610 & 280 \\    
13 &  27.0 & $-$27.5 &  26.9 & 1565 & 200 \\    
14 &  27.0 &  55.0 &   3.4 & 1567 & 229 \\    
15 &  27.0 & $-$55.0 &  27.4 & 1567 & 245 \\    
16 &  27.0 &  82.5 &   2.5 & 1553 & 289 \\    
17 &  27.0 & $-$82.5 &  14.0 & 1604 & 270 \\    
18 &  27.0 & 110.0 & \nodata\tablenotemark{d} &\nodata &\nodata \\
19 &  27.0 &$-$110.0 &   2.6 & 1698 & 166 \\    
20 &  27.0 &$-$137.5 &   1.2\tablenotemark{c} & 1618 & 308 \\    
21 & $-$27.0 &   0.0 &  14.4 & 1627 & 165 \\    
22 & $-$27.0 &  27.5 &   9.6 & 1609 & 144 \\    
23 & $-$27.0 & $-$27.5 &  13.0 & 1598 & 255 \\    
24 & $-$27.0 &  55.0 &   4.4 & 1582 & 150 \\
25 & $-$27.0 & $-$55.0 &   7.7 & 1621 & 279 \\   
26 & $-$27.0 &  82.5 &   2.7 & 1626 & 241 \\   
27 & $-$27.0 & $-$82.5 &   5.3 & 1658 & 321 \\   
28 & $-$27.0 & 110.0 &   2.2 & 1568 & 463 \\   
29 & $-$27.0 &$-$110.0 &   3.7 & 1640 & 302 \\   
30 & $-$27.0 &$-$137.5 & \nodata\tablenotemark{d} &\nodata &\nodata \\
31 &  53.9 &   0.0 &  11.8 & 1592 & 213 \\   
32 &  53.9 &  27.5 &   7.7 & 1572 & 246 \\   
33 &  53.9 & $-$27.5 &  18.9 & 1558 & 215 \\   
34 &  53.9 &  55.0 &   2.4 & 1648 & 204 \\   
35 &  53.9 & $-$55.0 &  19.8 & 1560 & 193 \\   
36 &  53.9 &  82.5 &   1.2 & 1652 & 241 \\   
37 &  53.9 & $-$82.5 &   9.1 & 1597 & 254 \\   
38 &  53.9 &$-$110.0 &   2.7 & 1630 & 360 \\   
39 &  53.9 &$-$137.5 &   1.5 & 1651 & 283 \\   
40 & $-$53.9 &   0.0 &   9.4 & 1628 & 241 \\   
41 & $-$53.9 &  27.5 &   3.3 & 1622 & 222 \\   
42 & $-$53.9 & $-$27.5 &   4.4 & 1669 & 245 \\   
43 & $-$53.9 &  55.0 &   2.2 & 1616 & 388 \\   
44 & $-$53.9 & $-$55.0 &   2.3 & 1696 &  62 \\    
45 & $-$53.9 &  82.5 &   1.1\tablenotemark{c} & 1547 & 150 \\   
46 & $-$53.9 & $-$82.5 &   0.9\tablenotemark{c} & 1788 &  25 \\    
47 & $-$53.9 &$-$110.0 &   2.2 & 1598 & 323 \\   
48 &  80.9 &   0.0 &   4.5 & 1607 & 316 \\   
49 &  80.9 &  27.5 &   2.3 & 1567 & 235 \\   
50 &  80.9 & $-$27.5 &   6.5 & 1567 & 228 \\   
51 &  80.9 &  55.0 &   0.6\tablenotemark{c} & 1734 & 147 \\    
52 &  80.9 &$-$137.5 &   1.2\tablenotemark{c} & 1725 & 127 \\   
53 & $-$80.9 &   0.0 &   1.7 & 1642 &  53 \\    
54 & $-$80.9 &  27.5 &   2.4 & 1650 & 303 \\   
55 & $-$80.9 & $-$27.5 &   1.6 & 1632 & 242 \\          
56 & $-$80.9 &  55.0 &   0.6\tablenotemark{c} & 1600 & 157 \\    
57 & $-$80.9 & $-$55.0 &   0.4\tablenotemark{c} & 1638 &  21 \\
58 & $-$80.9 & $-$82.5 &   0.5\tablenotemark{d} & 1571 &  74 \\    
59 & 109.3 &   0.0 &   1.2 & 1554 & 186 \\    
60 & 109.3 &  27.5 &   1.3\tablenotemark{c} & 1697 & 152 \\    
61 & 109.3 & $-$27.5 &   0.9\tablenotemark{c} & 1610 & 179 \\    
62 &$-$107.9 &   0.0 & \nodata\tablenotemark{d} &\nodata &\nodata \\    
63 &$-$107.9 &  27.5 &   1.0\tablenotemark{c} & 1502 & 100 \\    
64 &$-$107.9 & $-$27.5 &   0.4\tablenotemark{c} & 1599 &  54 \\    
65\tablenotemark{e} &  81.8 & $-$55.0 &   4.6 & 1538 & 213 \\  
66\tablenotemark{e} & 109.3 & $-$82.5 &   1.7 & 1584 & 178 \\   
67\tablenotemark{e} &  81.8 & $-$82.5 &   3.5 & 1575 & 284 \\   
68\tablenotemark{e} &  81.8 &$-$110.0 &   0.9 & 1629 & 251 \\   
69\tablenotemark{e} & 109.3 & $-$55.0 &   1.7 & 1564 & 201 \\   
70\tablenotemark{e} & 109.3 &$-$110.0 & 1.7 & 1617 & 324 \\   
71\tablenotemark{e} & 136.8 & $-$55.0 & 0.4\tablenotemark{c} & 1588 & 30 \\   
72\tablenotemark{e} & 136.8 & $-$82.5 & 1.2\tablenotemark{c} & 1644 & 370\\   
73\tablenotemark{e} & 136.8 &$-$110.0 & 0.5\tablenotemark{d} & 1520 & 300\\   
74 &  25.9 & $-$39.1 &  34.3 & 1558 & 209 \\   

\tablenotetext{a}{Offsets relative to RA=$12^h01^m53.1^s$, 
Dec=$-18^\circ 52^{'}05.^{''}0$ (J2000, the nucleus of NGC~4038).}    
\tablenotetext{b}{Integrated CO line intensity,
which is calculated using the same broad line emission 
window of 1300 to 1900 \kms. The Jy/K ($T_{\rm R}^*$) conversion 
adopted is $\sim 35$ Jy/K. Some offsets have slightly larger
$I_{\rm CO}$ when a narrower line emission window is used.}
\tablenotetext{c}{Tentative detections ($\approxgt 3 \sigma$).}    
\tablenotetext{d}{Non-detections ($< 3 \sigma$).}
\tablenotetext{e}{Deep CO spectra are shown in Figure~5a.}

\enddata

\end{deluxetable}

\clearpage

\begin{deluxetable}{llr}
\tablecaption{Global Properties of Arp~244 \label{tbl-2}}
\tablehead{
\colhead{Parameter}   &   \colhead{Value}   &   
\colhead{References}}

\startdata

 $d_{\rm L}$   & 20 Mpc & van der Hulst 1979; Mirabel \etal 1998 \\
 $L_{\rm B}$   & $2.9\times 10^{10} \ls$  & from B mag. in RC3 \\
 $L_{\rm IR}$  & $6.2\times 10^{10} \ls$ ($IRAS$) & Soifer \etal 1989; Sanders \etal 2001 in prep. \\
 $L_{\rm IR}$  & $8.0\times 10^{10} \ls$ (KAO) & Bushouse \etal 1998 \\
 $I_{\rm CO}$  & 90.6~K~\kms~ ($T_R^*$) & this work \\ 
 $S_{\rm CO}dV$  & 3172~Jy~\kms~ & this work \\ 
 $L_{\rm CO}$  & $0.3\times 10^{10}$~K~\kms~pc$^2$  ($T_{\rm mb}$) & this work \\
  M(H$_2$)     & 1.5$\times 10^{10} \ms$ & this work \\
 $L_{\rm IR}/$M(H$_2$)     & 4.2 \ls/\ms & this work \\
 M(HI)         & $0.4 \times 10^{10} \ms$ & van der Hulst 1979; Hibbard \etal 2001 in prep. \\
 M$_{\rm dust}$ & $\sim 10^8$ \ms & Haas \etal 2000 \\
 M$_{\rm gas}/$M$_{\rm dust}$  & $\sim 190 $ & this work \\
 $L_{\rm HCN}$ & $0.7\times 10^8$~K~\kms~pc$^2$ ($T_{\rm mb}$) & this work \\
 $L_{\rm HCN}/L_{\rm CO}$  & 0.02 & this work \\
 M(H$_2$)$_{\rm tidal~dwarf}$ & $\approxgt 2\times 10^8$~\ms & this work \\
\enddata

\end{deluxetable}

\clearpage

\begin{figure*}
{\Huge 0010128.fig1.gif}\\ 
\\
\epsscale{2.0}
Fig.~1. DSS image of Arp~244 (the ``Antennae'' galaxies)
and the locations of our deep HCN and CO search.
Thin circles are CO beams, observed at southeastern edge of the
disk overlap where the southern tidal tail begins, and at the
tidal dwarf galaxy near the tip of the southern tail. 
The thick circles mark the two HCN beams.
The southern HCN beam covers both the overlap and the nuclear region of
NGC~4039. 

\end{figure*}

\clearpage

\begin{figure*}
{\Huge 0010128.fig2.gif}\\
\\ 
\epsscale{1.9}
Fig.~2. Fully mapped NRAO 12m CO(1-0) integrated 
intensity map (in contours) overlaid on 
the HST/WFPC2 image. The contours are 3, 4, 6 to 28 K~\kms~ 
(in increases of 2 K~\kms~ on a $T_R^*$ scale). The CO 
spectra at each of the observed position are also plotted (velocity 
ranges from 1150 to 2050~\kms, and $T_R^*$~ ranges from $-25$ to 155 mK).
The currently accepted total molecular gas mass in the 
literature was based upon only one beam measurement
near the CO peak position (Sanders \& Mirabel 1985). Obviously one
NRAO 12m beam (top left) does not capture anywhere even near the CO emission
from the overlap region. We have sampled
73 positions at half-beam spacing to ensure that no further 
extended CO emission
was significantly detected in the outer edge, in typically 
1.5 hr of the integration time.
\end{figure*}

\clearpage

\begin{figure*}
\epsscale{2.0}
\plotone{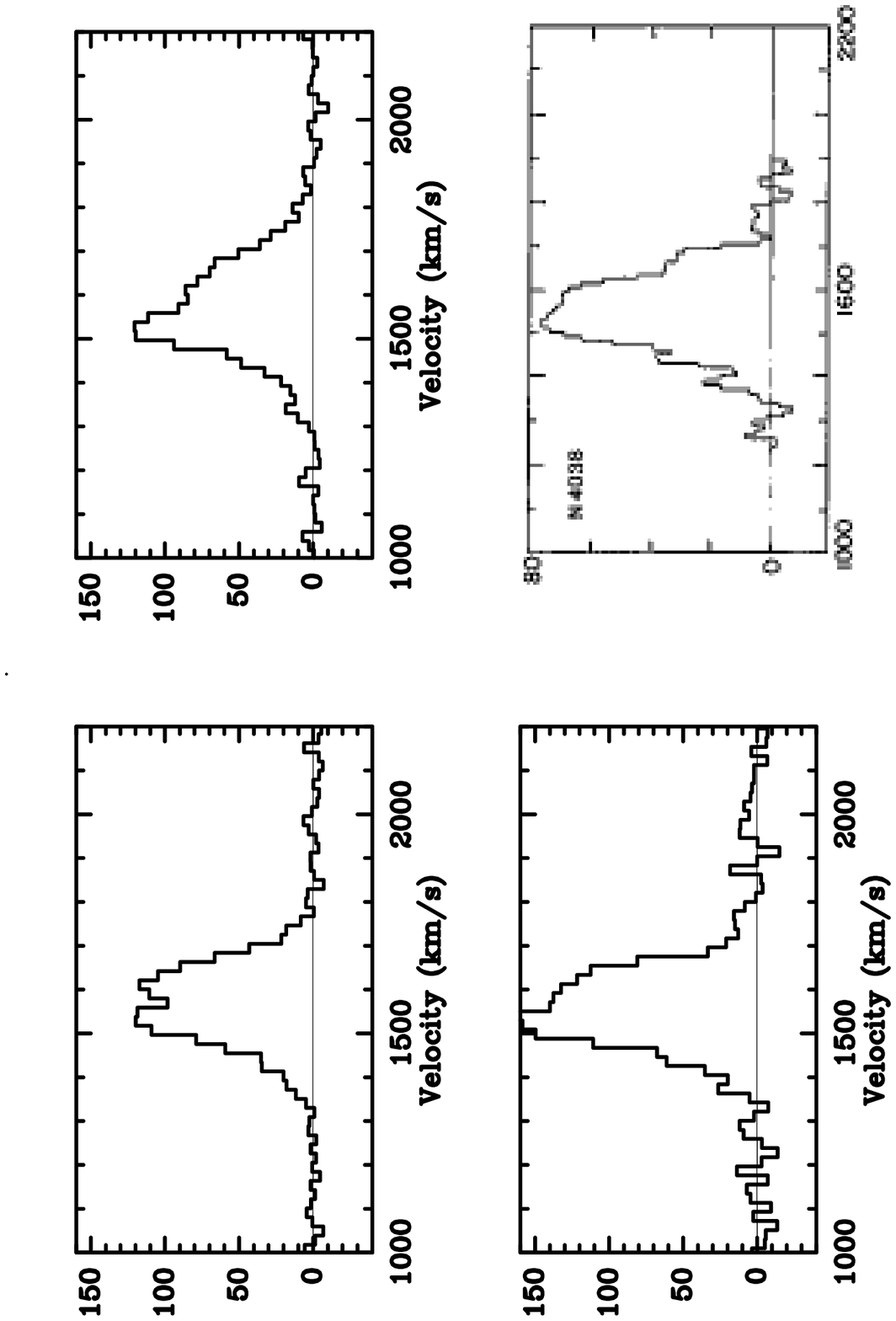}
Fig.~3. CO spectra closest to the peak CO emission in the
overlap region compared with the old CO spectrum of
Sanders \& Mirabel (1985). We detected broader CO
profile, which was previously missed due to
insufficient velocity coverage in the old
CO observation, and stronger CO emission.
\end{figure*}

\clearpage

\begin{figure*}
{\Huge 0010128.fig4a---4e.gif}\\
\\
 
\epsscale{2.}
Fig.~4a---4e. CO integrated intensity maps (contours) in 100~\kms velocity 
width. All contours plotted are 
0.4, 0.7 to 2.5 K~\kms~ in increases of 0.3 K~\kms. 
\end{figure*}

\clearpage









\begin{figure*}
{\Huge 0010128.fig4f.gif}\\
\\ 
\epsscale{2.}
Fig.~4f. The CO integrated intensity map over the  total 
velocity range (essentially same as the contour map in Figure~2, 
but here all CO data have been used),
which has contours 1.4, 2, 3, 4, 6 to 34 K~\kms, in increases of 2 K~\kms.
\end{figure*}

\clearpage

\begin{figure*}
\epsscale{1.8}
\plotone{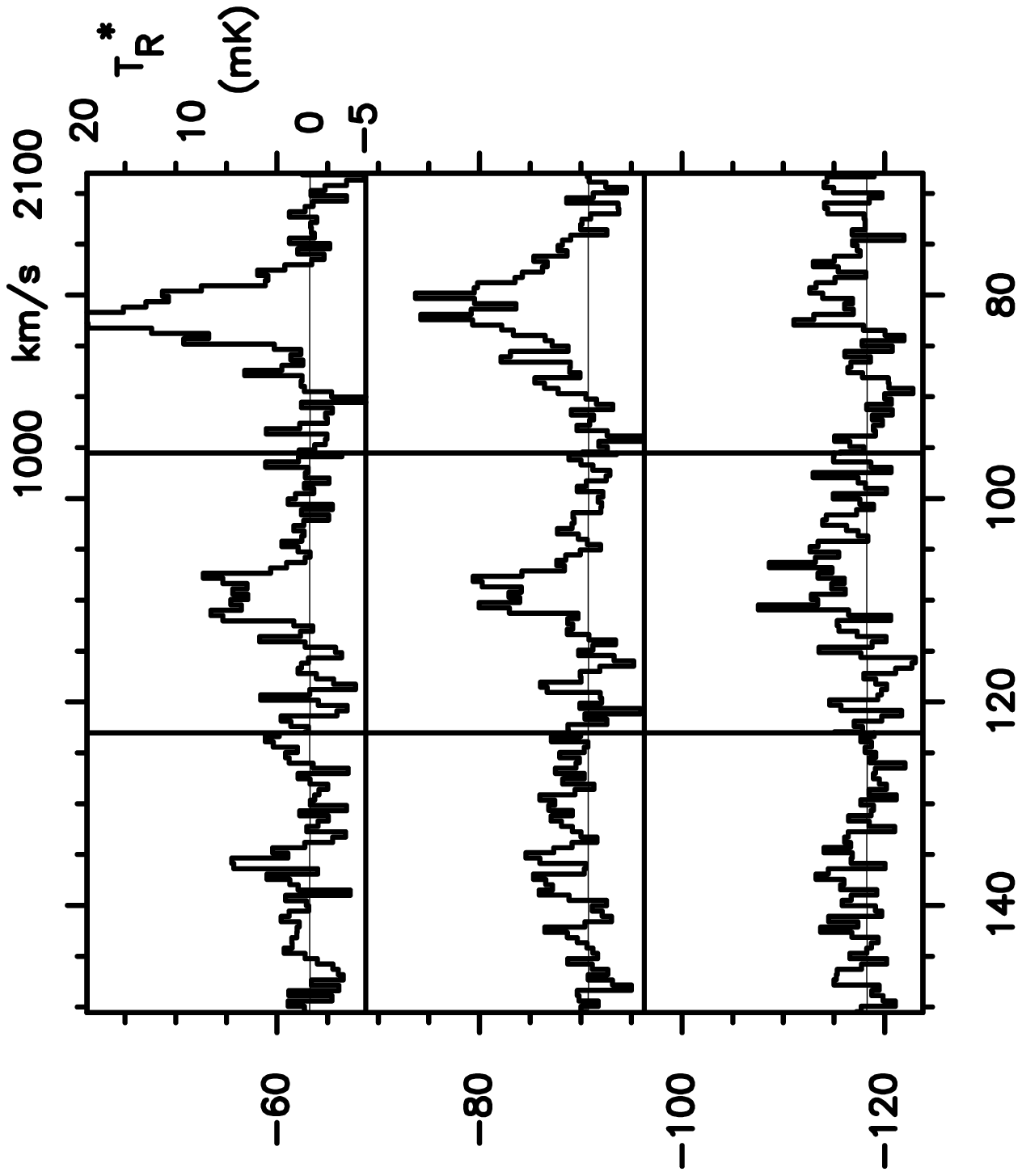}
Fig.~5a. The deep CO(1-0) spectra near the southeastern (SE)
edge of the overlap region (the SE corner of the merging disks
in Fig.~1).
The numbers refer to the offsets in arcsec, with respect to the 
northern nucleus, and the central position here 
corresponds to an offset of (110$''$, $-82.5''$). 
Fig.~5b. Both the summed average spectrum (top)
from all 5 positions searched for CO (more than 27 hours integration) 
in the tidal dwarf galaxy, and the deep integration spectrum (bottom)
at position of RA=12:01:26.0, DEC=$-$19:00:37.5 (J2000)
(more than 12 hour integration), 
show tentative weak detections. Fig.~5c. HCN(1-0) spectra obtained from 
the northern galaxy NGC~4038
and from the overlap region plus the southern galaxy NGC~4039.
\end{figure*}

\clearpage

\begin{figure}
\epsscale{1.75}
\plotone{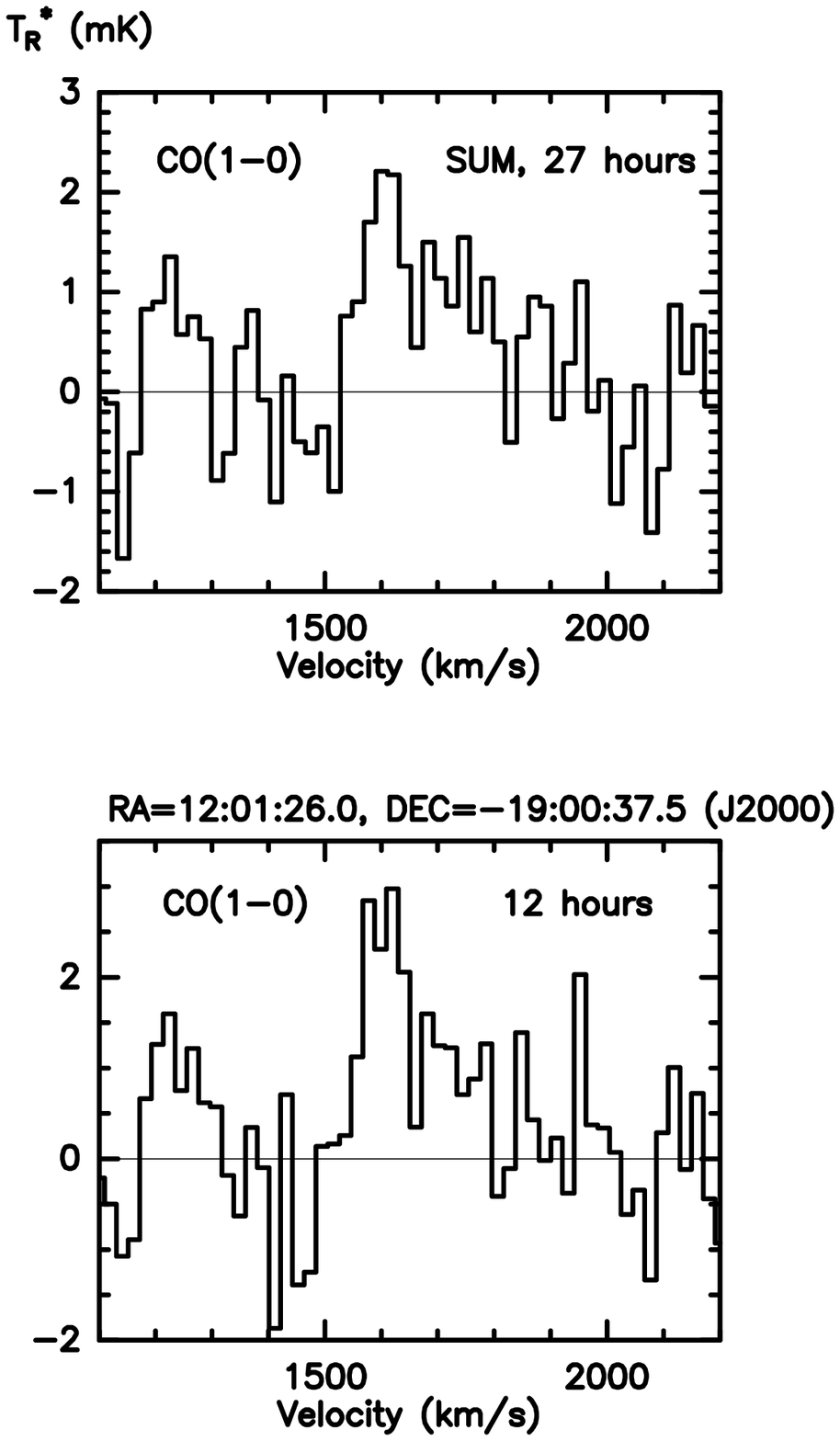}
Fig.~5b. 
\end{figure}
\begin{figure}
\plotone{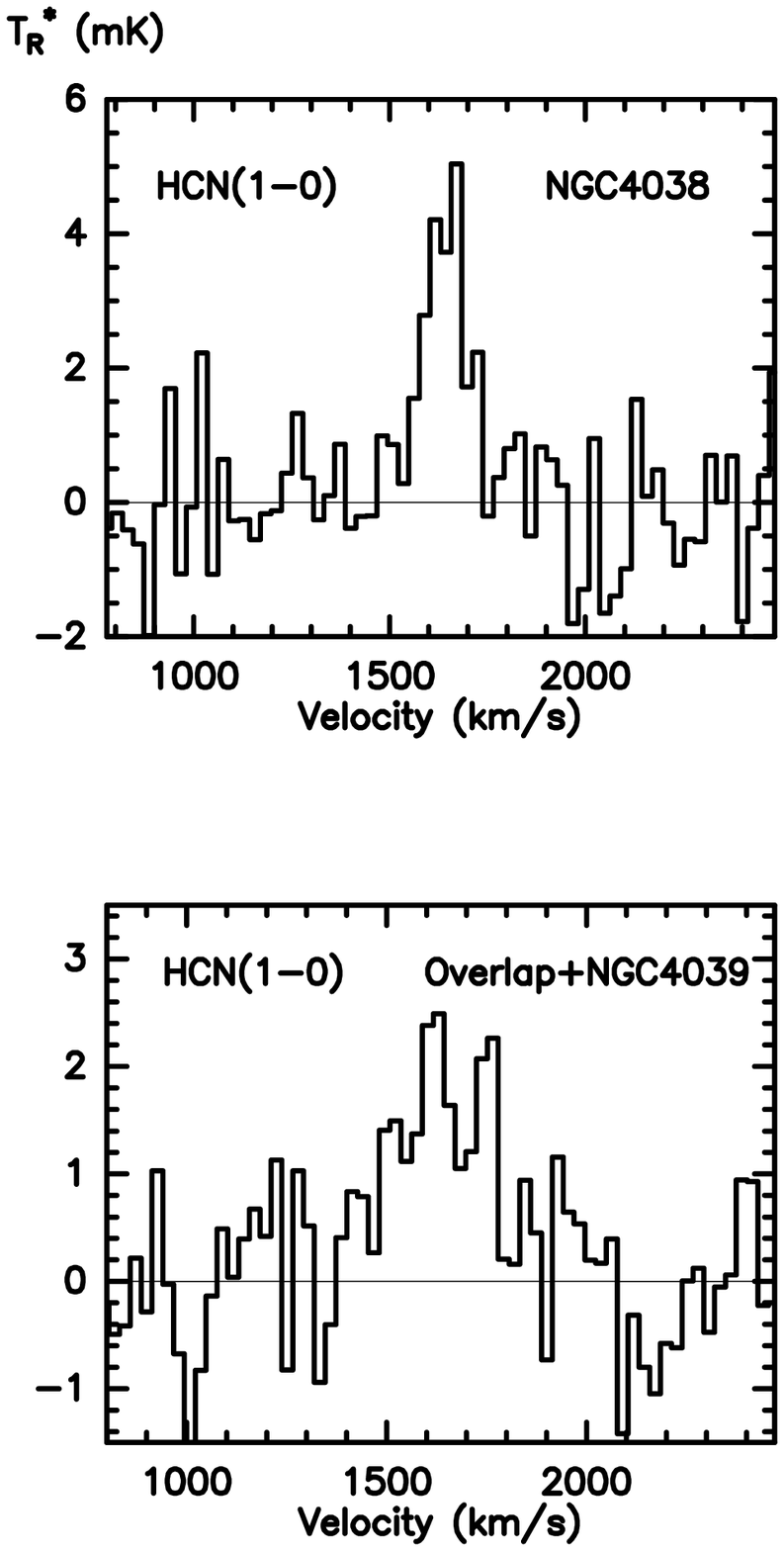}
Fig.~5c
\end{figure}

\clearpage

\begin{figure*}
{\Huge 0010128.fig6.gif}\\
\\ 
\epsscale{2.2}
Fig.~6. The 20cm radio continuum to CO(1-0) emission ratio 
map (contours) compared with  the HST/WFPC2
image (gray-scale). This illustrates the sites of the most intense starbursts
of the highest star formation efficiency (SFE). White contours indicate
the sites of the highest ratios, $\sim 5$ times above the average, 
suggesting the highest SFE $\approxgt 20~\ls/\ms$. Black contours are
the sites of a factor of 2 lower in the radio-to-CO ratios,
implying a SFE $\approxgt 10~\ls/\ms$.
\end{figure*}

\clearpage

\begin{figure*}
{\Huge 0010128.fig7.gif}\\
\\ 
Fig.~7. The overall ratio map (contours) of the 20cm radio continuum to 
the CO emission compared with the 
VLA 20cm radio continuum image (gray-scale). Here black contours are
the sites of the highest radio-to-CO ratios (same as those white 
contours in Figure~6) and the ones a factor of 1.5 lower,
gray contours are for the radio-to-CO ratios of a factor of 2 
lower than the highest (same as those black contours in Figure~6) 
and those a factor of 1.5 further lower, and white contours here are 
for the lowest radio-to-CO ratios, a factor of 4 and 6 lower than the 
highest, indicating the sites of SFE $\sim 5 \ls/\ms$ and lower,
about the average SFE across over the entire system.
\end{figure*}
\end{document}